\documentclass[aps,pre,twocolumn,citeautoscript,superscriptaddress,footinbib,eqsecnum]{revtex4-1}
\pdfoutput=1 
\usepackage{amsmath}
\usepackage{amssymb}
\usepackage{color}
\usepackage{mathtools}
\usepackage{graphicx}

\makeatletter

\usepackage{enumitem}
\usepackage{amsthm}
\usepackage{braket}
\usepackage{wasysym}
\usepackage{enumerate}
\usepackage{simplewick}
\usepackage{empheq}
\usepackage{float} 
\usepackage{multirow}
\usepackage{listings}
\usepackage{amsfonts}
\usepackage{bm}
\usepackage{mathrsfs}
\usepackage{empheq}
\usepackage{color}
\usepackage{setspace}
\usepackage{upgreek}
\def\({\left(}
\def\){\right)}
\def\weight[#1,#2,#3]{\{(#1),#2,#3\}}

\newcommand{\nn}{\\ \nonumber}

\renewcommand{\vec}[1]{\boldsymbol{#1}}

\newcommand{\avg}[1]{\left\langle #1 \right\rangle}
\usepackage{subfig}
\begin{document}
\title{A new role for circuit expansion for learning in neural networks}
\author{Julia Steinberg}
\email[Correspondence to: ]{jsteinberg@princeton.edu}
\affiliation{Center for Brain Science, Harvard University, Cambridge MA 02138,
USA}
\affiliation{Department of Physics, Harvard University, Cambridge MA 02138, USA}
\author{Madhu Advani}
\affiliation{Center for Brain Science, Harvard University, Cambridge MA 02138,
USA}
\author{Haim Sompolinsky}
\affiliation{Center for Brain Science, Harvard University, Cambridge MA 02138,
USA}
\affiliation{Edmond and Lily Safra Center for Brain Sciences, Hebrew University,
Jerusalem 91904, Israel }
\date{\today}
\begin{abstract}
Many sensory pathways in the brain include sparsely active populations
of neurons downstream from the input stimuli. The biological purpose of this expanded structure is unclear,
but may be beneficial due to the increased expressive power the  
network. In this work, we show that certain ways of expanding a neural network can improve its generalization performance even when
the expanded structure is pruned after the learning period. To study
this setting, we use a teacher-student framework where a perceptron
teacher network generates labels corrupted with small amounts
of noise. We then train a student network structurally matched
to the teacher. In this scenario, the student can achieve optimal accuracy if given the teacher’s
synaptic weights. We find that sparse expansion of the input layer of a
student perceptron network both increases its capacity and improves
the generalization performance of the network when learning a noisy
rule from a teacher perceptron when the expansion is pruned after
learning. We find similar behavior when the expanded units are stochastic
and uncorrelated with the input and analyze this network in the mean
field limit. By solving the mean field equations, we show that the
generalization error of the stochastic expanded student network continues
to drop as the size of the network increases. This improvement in generalization
performance occurs despite the increased complexity of the student
network relative to the teacher it is trying to learn. We show that
this effect is closely related to the addition of slack variables
in artificial neural networks and suggest possible implications for
artificial and biological neural networks. 
\end{abstract}
\maketitle

\section{Introduction}
Learning and memory is thought to occur mainly through long term modification of synaptic connections among neurons, a phenomenon well established experimentally. Additionally, neural circuits also undergo structural changes on a global level. It is observed that synaptic density in the human cortex increases rapidly after birth and then drops sharply towards adulthood, indicating an extensive pruning of the neuronal circuits \cite{Sakai2020}. Another form of structural plasticity, also occurring in the adult brain, is the continuous recycling of synapses which is seen in both cortex and hippocampus. In the past, several modeling studies have addressed the computational consequences of these phenomena (see \cite{neurogenesis2,neurogenesis3} for adult neurogenesis and \cite{neurogenesis4} for synaptic recycling). 

In this work, we explore a novel computational benefit of structural dynamics in neural circuits that learn new associations or tasks. We show that under certain classes of learning paradigms, the expansion of a neural circuit architecture by recruiting additional neurons and synapses may facilitate the dynamics of learning. Expanding circuit sizes to enable sparse coding has been shown to have computational benefits in several contexts of neuroscience and machine learning for sensory processing, learning and memory \cite{olshausensparse,doi:10.1146/annurev-neuro-062111-150410,kumar,doi:10.1002/hipo.450040319,Tsodyks_1988}. In these models, circuit expansion and the resultant sparse coding yield better representations of the stimuli, enhancing pattern separation, and improving the capacity for pattern retrieval and classification. Importantly, to realize these benefits, the expanded architecture needs to be stable after learning. By contrast, in our scenario, the benefit of expansion lies in its facilitating the dynamics of learning and not its information bearing potential. In fact, expansion in this scenario is most beneficial when it is transient, i.e. the added neurons and synapses are pruned after the learning period. Hence this hypothesis is consistent with the observed continuous recycling of synapses during learning.

\smallskip{}
Within this work, we consider neural networks that learn supervised classification problems
implemented by a single layer perceptron. Despite the apparent simplicity of this task, learning the rule
by training with labeled examples may be hampered by the complexity
of the underlying data. We focus on two cases of unrealizable
rules, which are characterized by a critical size of the
training set above which no single layer student is able to correctly
classify all of the training examples. This critical size is called
the student's capacity. We first consider unrealizable rules occurring when the teacher network produces training
labels corrupted by stochastic noise and will later consider cases in which 
the teacher is more complex than the student network trying to learn
the rule. We show that adding sparse expansions to student
networks by random mappings of the original input increases the capacity
of the student network and improves the generalization performance of the
network as it is trained on larger training sets. 

While the capacity of a network is clearly related to its dimensionality,
it is not obvious and even counterintuitive that increasing the size
of a network should improve its generalization performance. Using
mean field theory and simulations of a wide range of network parameters,
we show that expansion of the architecture during learning achieves
improved generalization, particularly if the additional elements of
the circuit are removed after learning. In addition, it is shown that
the effect is more pronounced if the hidden representation during
learning is sparse. We find that the performance is most improved
when the expanded units are random and uncorrelated with the original
input which suggests that having low overlap in expanded activity
between different training input is crucial to improving performance. 

\smallskip{}
Our analysis offers a new perspective on the important issue of the
relation between model complexity and learning in neural networks.
Artificial neural networks have achieved state of the art predictive
performance on a variety of tasks, especially within the past decade \cite{lecun2015deep,schmidhuber2015deep}.
The primary benefit to training these enormous models appears to lie
in their ability to represent very complex functions and the link
between width, depth, and expressivity of neural networks is discussed
in detail in several studies including, \cite{bengio2011expressive,poole2016exponential,Safran:2017:DTA:3305890.3305989,pmlr-v70-raghu17a}.
These networks are often over-parameterized in the sense that than
the number of examples the network is trained on is far less than
the number of free parameters in the network \cite{simonyan2014very}.
Classical statistical learning theory suggests that such massively
over-parameterized models should be expected to over-fit on the training
data \cite{zhang2016understanding} and make poor predictions on
new inputs not seen by the network before. To resolve this apparent
paradox, it has been suggested that modern learning algorithms cost
functions, and architectures incorporate strong explicit and implicit
regularizations \cite{bartlett2002rademacher,Advani2017a,Advani2017b,foundationsofmachinelearning}.
Our findings suggest there may be advantages to making neural networks
larger than is required for expressing the underlying task. These
advantages are related to enhancing the ease of the learning convergence,
and that in these cases, optimal performance after learning is achieved
upon removal of the additional nodes and weights. Indeed, pruning
of Deep Neural Networks after training is a current topic of research
in machine learning \cite{NIPS2015_5784,8100126,pruningreview,stateofsparsity}. 

\smallskip{}

We start in section \ref{sec:sparseexpansion} by showing in simulations
that implementing a sparse expansion of a perceptron network via random
mapping of the input can improve its generalization ability when learning
from a noisy teacher. In section \ref{sec:perceptrontheory} we analyze
these results by studying a simpler model of a single layer perceptron
in which the activity in the expanded units is random and uncorrelated
with the stimulus. We use the replica method to derive a mean field
theory exact in the thermodynamic limit, and find it matches well
with simulations of large but finite size networks. In section \ref{sec:slack}
we explain this phenomena more intuitively by showing a correspondence
between adding random input neurons and including \emph{slack variables}
in the optimization problem. We also discuss how hidden units in our
two layer network model can resemble the stochastic expansion of the
input layer in the one layer model. In section \ref{sec:mismatch}
we demonstrate how the benefit of sparse expansion also applies in
more general cases of learning unrealizable rules by comparing the
performance of a student learning from a more complex teacher network to
our theory results. In most of our work we have focused on convex
learning algorithms. In section \ref{sec:otheropt} we discuss to
what extent these effects extend to other learning algorithms. Finally,
we close by discussing some general implications of our results.
\section{Sparse Expansions and Learning}
\label{sec:sparseexpansion} 
\begin{figure}[h]
\includegraphics[width=0.45\textwidth]{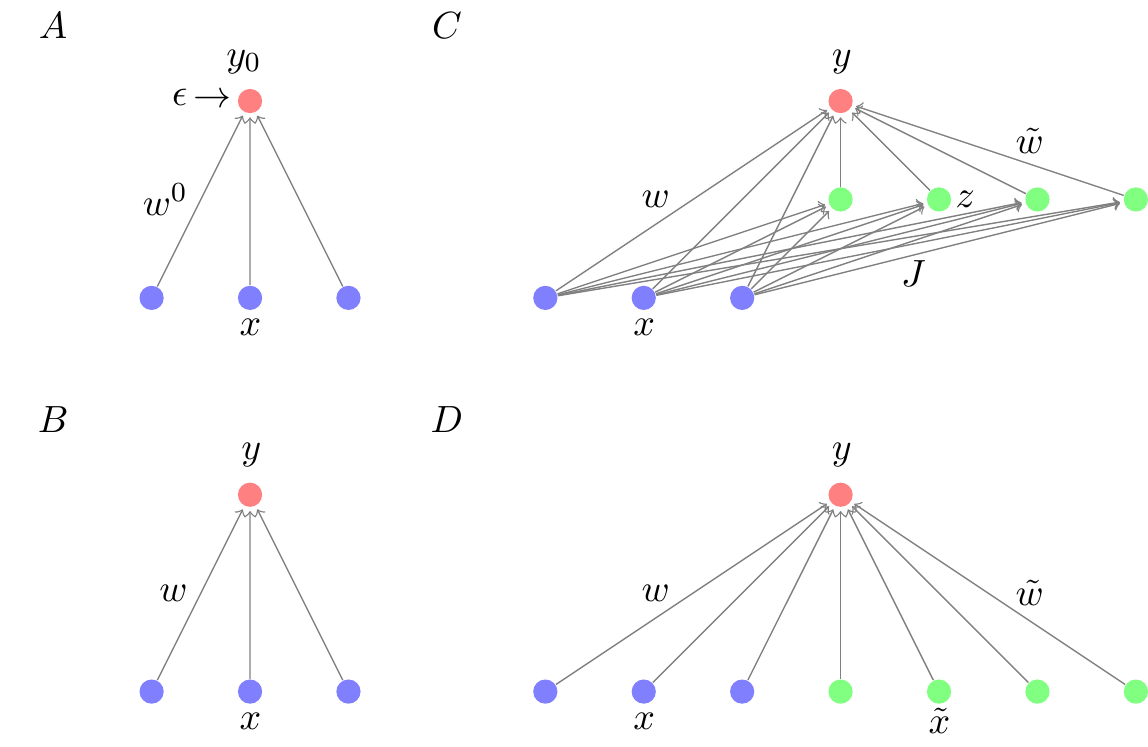}
\captionsetup{justification=raggedright,
singlelinecheck=false
}\caption{\label{fig:Teacher-and-student}Teacher and student network schematics.
(A) noisy teacher network (B) student network (C) Student with sparse
hidden layer (D) Student with stochastic expanded units.}
\end{figure}
We begin our analysis by considering a teacher perceptron network
with $N_{0}$ input nodes $x_{i}$, one output node $y_{0}$, and
$N_{0}$ synaptic weights $w_{i}^{0}$ drawn iid from $w_{i}^{0}\sim\mathcal{N}\left(0,\sigma_{w}^{2}\right)$
and supervised learning tasks in which a student perceptron will attempt
to learn the teacher's input-output rule from a training set provided
by it. For each input $x$ drawn iid from $x_{i}\sim\mathcal{N}\left(0,1\right)$,
the teacher network assigns a label $y_{0}\in\lbrace-1,1\rbrace$
via the following rule: $y_{0}=\text{sign(\ensuremath{h_{0}})}$ where the teacher field $h_{0}$ is given by

\begin{align}
h_{0} & =\frac{1}{\sqrt{\sigma_{w}^{2}N_{0}}}\sum_{i=1}^{N_{0}}{w_{i}^{0}x_{i}}+\epsilon\label{eq:trueteacher}
\end{align}
and $\epsilon\sim\mathcal{N}(0,\sigma_{out}^{2})$ denotes an output
or label noise (Fig.\ \ref{fig:Teacher-and-student} A). We assume
a training set consisting of $P$ such input-output pairs, and we
define $\alpha_{0}=P/N_{0}$ as the measurement density of training
examples relative to the teacher. 

The goal of training is to yield network weights that perform well
on new inputs, i.e., to have a small generalization error $E_{g}$,
defined as the expected fraction of mislabeled examples averaged over
the full distributions of inputs $x$ and the noise $\epsilon$ as
follows

\begin{eqnarray}
E_{g}(w) & = & \left\langle \Theta\left(-y_{0}(x)y(x)\right)\right\rangle _{x,\epsilon}
\end{eqnarray}
where $\Theta(x)$ is the Heaviside step function and the student labels $y(x)$ are given by 
\begin{equation}
y=\text{sign}\left(\frac{1}{\sqrt{N_{0}}}\sum_{i}^{N_{0}}{w_{i}x_{i}}\right)
\end{equation}
The generalization error is minimized when the student weights equals
those of the teacher, i.e. $w=w_{0}$ . This will yield the same generalization
error as the teacher itself if it were tested on examples with labels
generated via Eqn.\ \ref{eq:trueteacher}. We refer to this error
as the minimal generalization error which can be expressed in terms
of the noise as follows

\begin{align}
E_{\text{min}} & =E_{g}(w^{0})=\frac{1}{\pi}\left(\frac{\pi}{2}-\tan^{-1}\left(\frac{1}{\sigma_{out}}\right)\right)\label{eq:oracle}
\end{align}
which provides a lower bound on the generalization error of a student
as no network architecture (even more complex than a perceptron) can
yield a better performance. 

Finding the optimal set of weights may be difficult even if the number
of examples is large. Due to label noise from the teacher, training
examples will no longer be linearly separable i.e., perfectly classified
by a perceptron, beyond some critical value of $P$, rendering the
training task as ``unrealizable'' by a perceptron. Furthermore,
unlike the realizable regime, in the unrealizable regime, finding
the minimum of the training error is a nonconvex problem and can be
hampered by local minima. Here we assume that the training is restricted
to minimizing the training error by applying convex algorithms. Such
training algorithms are limited to sizes smaller than the capacity.
The capacity depends on the level of output noise in the labels, and is shown as a function of $\sigma_{out}$ in
Fig.\ \ref{fig:capacity}.

For a teacher of fixed width $N_{0}$ and a fixed training set of
size $P$, we can increase the capacity of the student network by
making the student network larger than the teacher.
There are several ways to expand the student network and each have
a different effect on the generalization performance.We first increase the network size by implementing a random transformation
of input stimuli to a hidden layer of size $N_{+}$ as depicted in
C of Fig.$\,$\ref{fig:Teacher-and-student}. The input of the full network is now $N=N_{0}+N_{+}$. The labels in the student
network are given by $y^{\mu}=\text{sign}(h^{\mu})$ where,
\begin{equation}
h^{\mu}=\frac{1}{\sqrt{N}}\left(\sum_{i=1}^{N_{0}}{w_{i}x_{i}^{\mu}}+\sum_{j=1}^{N_{+}}{\tilde{w}_{j}z_{j}^{\mu}}\right)\label{eq:deterministicstudent}
\end{equation}
where $z^{\mu}$ represents the activity in a hidden layer of neurons
generated by a random connectivity matrix $J$ , 

\begin{equation}
z_{j}^{\mu}=\frac{A}{\sqrt{f(1-f)}}\left(\Theta\left(\sum_{i=1}^{N_{0}}J_{ji}x_{i}^{\mu}-T\right)-f\right)\label{eq:twolayerhidden}
\end{equation}

where $A$ is a positive scalar and $T$ is a firing threshold chosen
to produce hidden layer neuronal activity with a given sparsity $f$.
The synapses $J_{ji}$ are chosen iid according to $J_{ji}\sim\mathcal{\mathcal{N}}\left(0,1\right)$
and are uncorrelated with the teacher network,

In simulations, we measure the performance of this network by estimating
the generalization error on new examples generated from the same distribution
as the training set ($x^{\mu},y_{0}^{\mu}$). Because $J$ is fixed,
the training problem is still that of linear classification with
an expanded input layer of size $N=N_{0}+N_{+}=\beta N_{0}$ and correspondingly
an expanded trained weight vector $(w,\tilde{w})$ . We train the
output weights using max-margin classification (i.e., Linear SVM \cite{vapniksvm,svmvapnik})
which finds an error free solution that maximizes the minimal distance
of the input examples from the separating plane, called margin, $\kappa$
, which in our case is defined through the linear inequality 
\begin{equation}
y_{0}^{\mu}h^{\mu}\geq\kappa||w+\tilde{w}||,\forall\mu\label{eq:kappa}
\end{equation}
provided that such a solution exists. Max-margin classification
is equivalent to solving the following quadratic programming problem with linear constraints, 
\begin{eqnarray}
(w^{*},\tilde{w}^{*}) & = & \arg\min_{w,\tilde{w}}{\sum_{i=1}^{N_{0}}w_{i}^{2}+\sum_{j=1}^{N_{+}}\tilde{w}_{j}^{2}}\label{eq:maxclass1}\\
 & s.t. & \quad y^{\mu}h^{\mu}\geq1\quad\forall\mu\label{eq:maxmarginclassifier}
\end{eqnarray}
The optimization problem in Eqns.\ \ref{eq:maxclass1}, \ref{eq:maxmarginclassifier} is
convex and admits a unique solution $(w^{\ast},\tilde{w}^{\ast})$.
We choose the max-margin solution as in general it is known to yield
a robust solution to the classification problem with good generalization
performance \cite{bartlettsvm,vapnik}.

\begin{figure}[h]
	\includegraphics[width=0.4\textwidth]{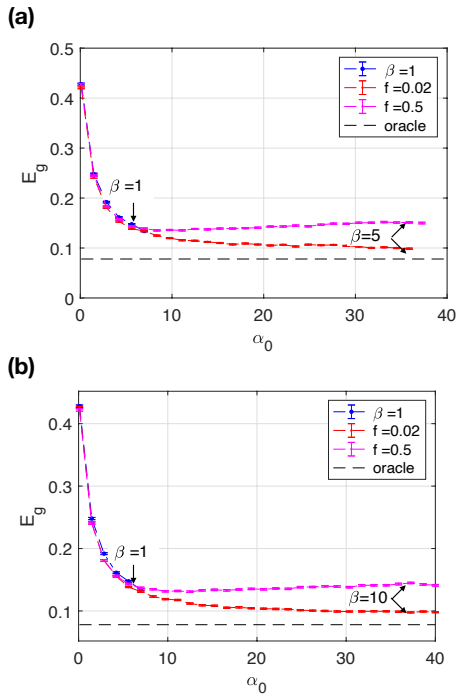}
	\captionsetup{justification=raggedright,
		singlelinecheck=false
	}\caption{\label{fig:spareexpansion}The generalization error $E_{g}$ from
	simulations of a two layer network after pruning the expanded units. (a) compares $E_{g}$
	as a function of $\alpha_{0}$ for a student network the same size
	as the teacher and for student networks with expansion factors $\beta=5$
	with dense ($f=0.5$) and sparse ($f=0.02$) activity. (b)
	does the same for $\beta=10$. The oracle line represents the lowest
	possible generalization error due to the presence of label noise.
	The parameters $A=0.2$, $\sigma_{out}=0.25$, and $N_{0}=100$ and
	$200$ trials are used in both figures.}
\end{figure}

\begin{figure}[h]
	\includegraphics[width=0.35\textwidth]{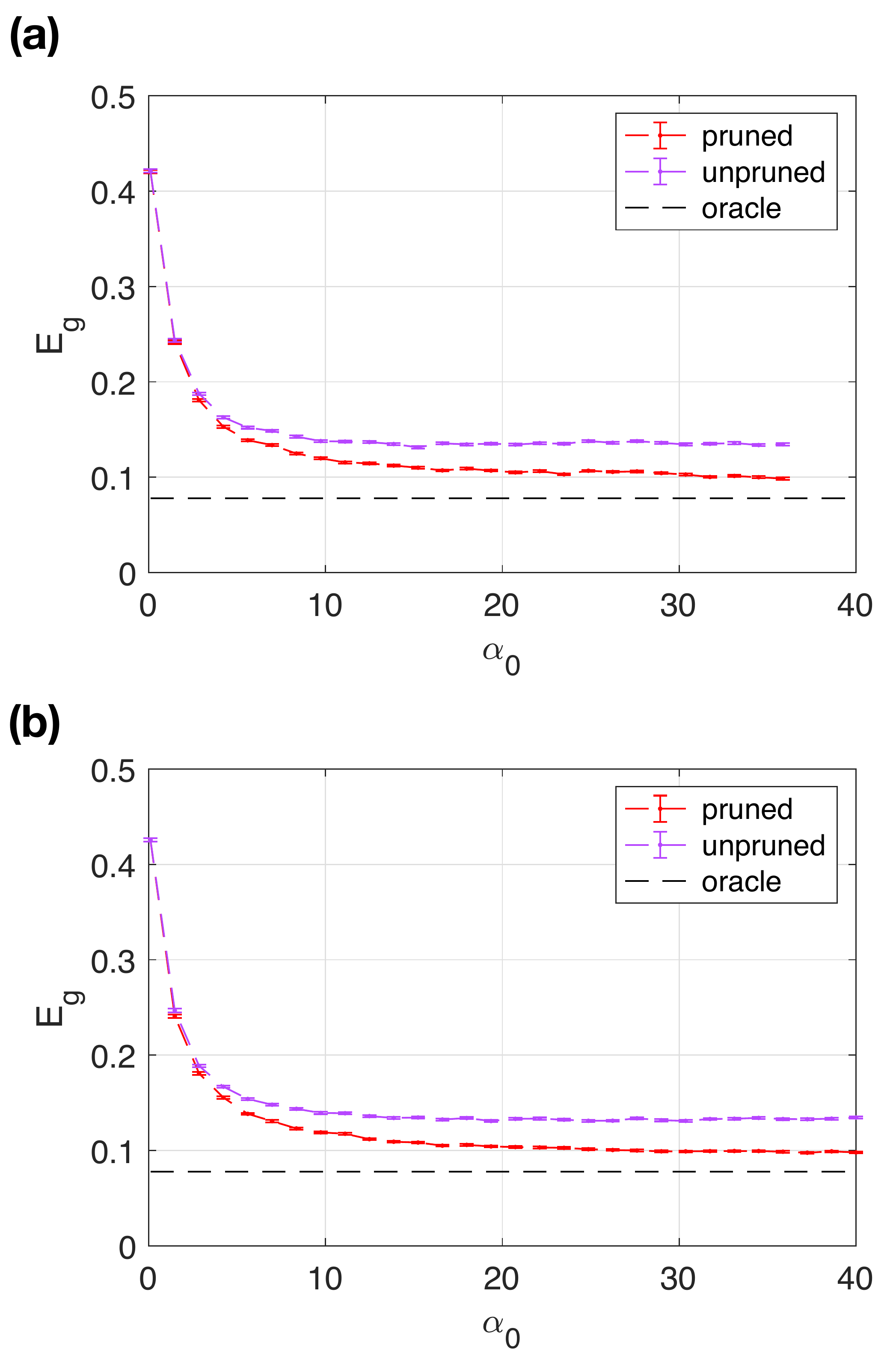}
	\captionsetup{justification=raggedright,
		singlelinecheck=false
	}\caption{\label{fig:sparseprunedunpruned}Comparison of the generalization
	error in simulations of a sparsely expanded two layer network before
	and after pruning the expanded units. (a) shows simulations
	for a student network with expansion factor $\beta=5$ and (b)
	shows simulations for a student network with $\beta=10$. We see for
	both values of $\beta$ the student network with the best overall
	performance is the network with sparse expansion with expanded weights
	are pruned after learning. The parameters $f=0.02$, $A=0.2$, $\sigma_{out}=0.25$,
	and $N_{0}=100$ and $200$ trials are used in both figures.}
\end{figure}

\begin{figure}[h]
	\includegraphics[width=0.4\textwidth]{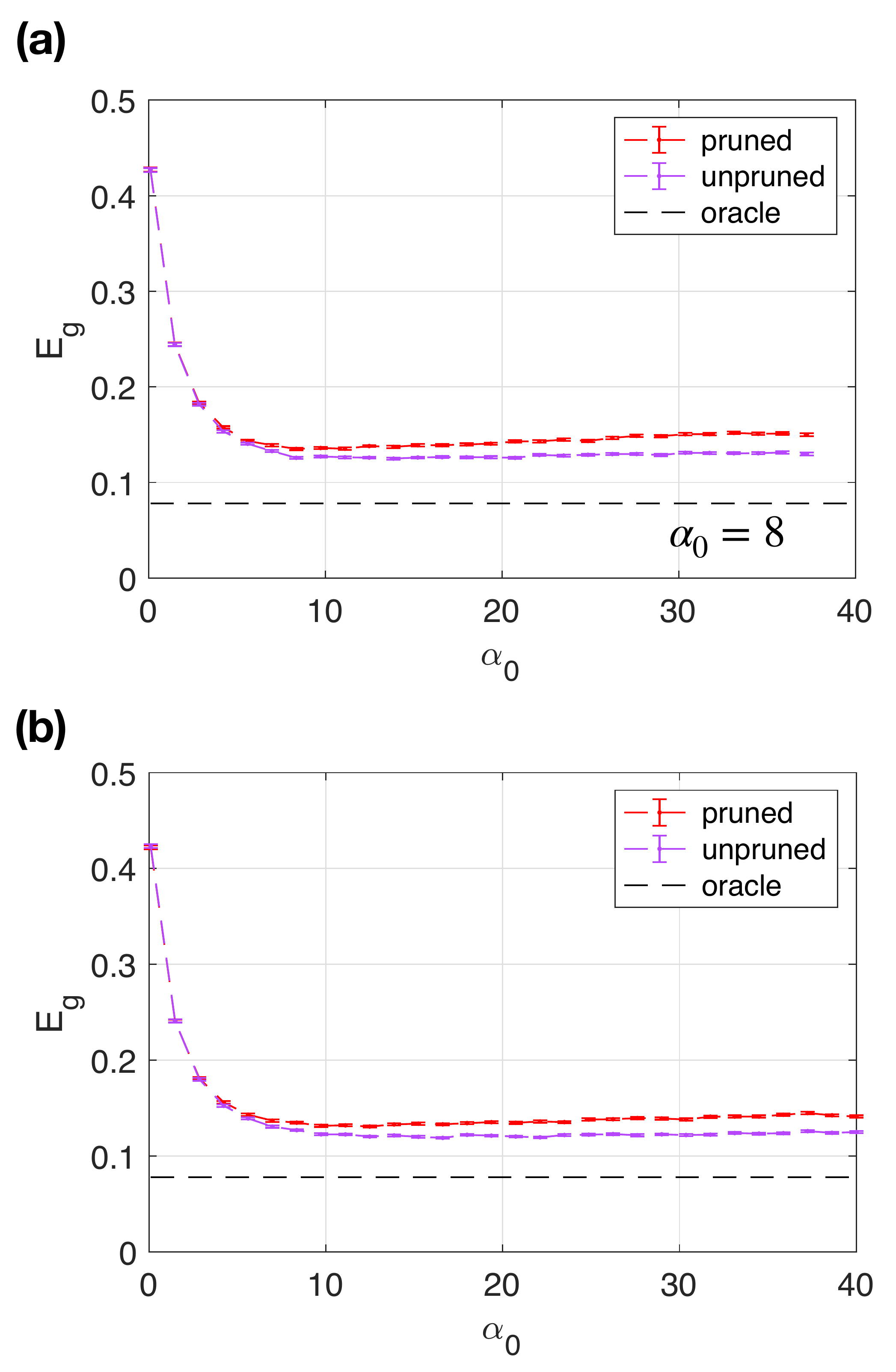}
	\captionsetup{justification=raggedright,
		singlelinecheck=false
	}\caption{\label{fig:denseprunedunpruned}Comparison of the generalization error
	from simulations of a densely expanded two layer network before and
after pruning the expanded units. (a) shows simulations
	for a student network with expansion factor $\beta=5$ and (b)
	shows simulations for a student network with $\beta=10$. We see that
	the densely expanded network performs best when the expanded weights
	are unpruned. However, the performance of the sparsely expanded network
	with pruned weights in \ref{fig:sparseprunedunpruned} is superior
	to the densely expanded network regardless of whether the weights
	are pruned or kept. The parameters $f=0.5$, $A=0.2$, $\sigma_{out}=0.25$,
	and $N_{0}=100$ and $200$ trials were used for all figures.}
\end{figure}

As expected, the addition of this hidden layer increases the capacity
of the student, namely the maximal value of $P$ for which the training
data are linearly separable \cite{4038449}. For instance, for the
parameters of Figs.\ \ref{fig:spareexpansion} (a) and \ref{fig:spareexpansion} (b), the
capacity increases from a maximum value of $\alpha_{0}$, equaling
$\sim6$ for no expansion $(\beta=1)$ to $\alpha_{c}\sim35$ and
$\sim75$ for $\beta=5$ and $10$, respectively. In limit $N_{0}\rightarrow\infty$,
it appears that this increased capacity does not depend on the sparsity
of the hidden layer, or depends on it very weakly. By
enabling the network to train successfully on a large training set
adding the random layer substantially improves the generalization performance
of the network, particularly, if the hidden layer activity $z^{\mu}$
is very sparse, i.e., $f\ll1$. As seen in Fig.\ \ref{fig:sparseprunedunpruned},
the generalization error decreases monotonically with increasing the
number of examples, up to the capacity. Furthermore, the generalization
performance of the network improves upon removal of the additional
neurons $N_{+}$ after learning. By contrast, for a hidden layer with
dense activity, the generalization error decreases initially with increasing
$\alpha_{0}$ but then saturates at an intermediate value of $\alpha_{0}$
and increases for larger values. For dense activity
the performance slightly deteriorates if the extra neurons
are removed after learning, shown in Fig.\ \ref{fig:denseprunedunpruned}. 

The role of sparseness will be discussed more thoroughly in section \ref{subsec:Correspondence-with-slack}.

%

The network represented in Eqns.\ \ref{eq:deterministicstudent} and \ref{eq:twolayerhidden}
is difficult to study analytically because of correlations in the
activities of the hidden layer induced by $J$ \cite{babadi2014sparseness}. We therefore consider in the following section a simplified expansion
scheme which we call, a \emph{stochastic} architecture, and is shown
in Fig. 1 D. In contrast to the \emph{deterministic} scheme of Fig.
1 C, the activity patterns of the additional neurons in this architecture are not
generated through connections from the input layer. Instead they are
randomly generated for each training pattern, $\mu$, independent
of $x_{\mu}$. The advantage of this scheme is that the random activities
of the hidden neurons are statistically independent of each other
and additionally for different training patterns, rendering the model amenable
to study using the tools of statistical mechanics. Although this scheme
is artificial from a biological perspective, we will show that when
the deterministic layer is very sparse the system's behavior is similar
to the stochastic model. 
\section{Theory of perceptron learning with expanded stochastic units}
\label{sec:perceptrontheory} In this section, we develop intuition
for the effect of sparse expansion on the generalization performance
of a perceptron by considering a simpler single layer student network
which can be solved analytically in the mean field limit. This network
(shown in B of Fig.\ \ref{fig:Teacher-and-student}) is trained using
data with $\mu=1,...,P$ binary labels $y^{\mu}$, generated by the
noisy teacher network in Eqn.\ \ref{eq:trueteacher}. \textcolor{black}{For
convenience, we keep the same normalization for the student and teacher
weight vectors which corresponds to setting $\sigma_{w}^{2}=\beta$}\textcolor{red}{{}
}in \ref{eq:trueteacher}. The activity of the student network takes
the form: 
\begin{equation}
h=\frac{1}{\sqrt{N}}\left(\sum_{i=1}^{N_{0}}{w_{i}x_{i}}+\sum_{j=1}^{N_{+}}{\tilde{w}_{j}\tilde{x}_{j}}\right)\label{eq:student}
\end{equation}
where $\tilde{x}_{j}^{\mu}$ are random units added to the input layer
and are drawn iid from a gaussian distribution with zero mean and
variance $\sigma_{in}^{2}$. The label $y$ given to input $x$ by
the student is $y(x)=\text{sign}(h)$. The student weights are trained
to yield the max margin solution in Eqn. \ref{eq:maxmarginclassifier}.
\subsection{Mean field theory}
We now analyze the performance of the expanded student network in \ref{eq:student}.
We will denote the measurement density of the training set relative
to the width of this student as $\alpha=\alpha_{0}/\beta$. The mean
field theory below is exact in the thermodynamic limit, where $P,N\rightarrow\infty$
and $\alpha\sim O(1)$ .

To perform an ensemble average of the system's properties over different
realizations of training sets, we use the replica trick in a manner
similar to \cite{Gardner_1987,Gardner_1988,Gardner_1988_2,seung1992statistical,RevModPhys.65.499,engel_van}. Full details of the
replica calculation and the form of the saddle equations are given
in Appendix \ref{app:replica}. 
We start by considering the version space for $n$ replicated students
indexed by $a$: 
\begin{eqnarray}
\langle V^{n}\rangle & = & \int\prod_{a}{dw^{a}d\tilde{w}^{a}\delta\left(\sum_{i=1}^{N_{0}}(w_{i}^{a})^{2}+\sum_{j=1}^{N_{+}}(\tilde{w}_{j}^{a})^{2}-N\right)}\nonumber \\
 & \times & \prod_{\mu=1}^{P}\sum_{\sigma=\pm1}\left\langle \Theta\left(\left[\sigma h^{\mu a}-\kappa\right]\right)\Theta(\sigma h_{0}^{\mu})\right\rangle \label{eq:V-1-2-1-1}
\end{eqnarray} 
where we have normalized the weights so that $\|w^{a}\|^{2}+\|\widetilde{w}^{a}\|^{2}=N$
in all replicas, and $\Theta$ is the Heavyside step function. The
quantities $h^{\mu a}$ are the student's fields induced by the $\mu$
-th input and weight vector $(w^{a},\tilde{w}^{a})$ of the $a$-th
replica; $h_{0}^{\mu}$ are the teacher fields induced by the $\mu$-th
input including noise. The angular brackets denote averaging with
respect to the gaussian input vectors, $x_{\mu}$(with variance $1)$,
student input noise vector,$\tilde{x}_{\mu}$ (with variance $\sigma_{in}^{2}$)
, teacher label noise, $\epsilon_{\mu}$ (with variance $\sigma_{out}^{2}$).
Since the distribution of inputs is isotropic, one
does not need to average over the teacher distribution. Evaluating
Eqn.\ \ref{eq:V-1-2-1-1}, we derive a mean field theory in terms of the
order parameters $m_{a}$, $\tilde{r}_{a}$, $q_{ab}$ and
$\tilde{q}_{ab}$ defined as
\begin{align}
m_{a} & =\frac{1}{N}\sum_{i=1}^{N_{0}}w_{i}^{0}w_{i}^{a}\label{eq:m}\\
\tilde{r}_{a} & =\frac{1}{N}\sum_{i=1}^{N_{+}}(\tilde{w}_{i}^{a})^{2}\label{eq:rt}\\
q_{ab} & =\frac{1}{N}\sum_{i=1}^{N_{0}}w_{i}^{a}w_{i}^{b}\label{eq:q}\\
\tilde{q}_{ab} & =\frac{\sigma_{in}^{2}}{N}\sum_{i=1}^{N_{0}}\tilde{w}_{i}^{a}\tilde{w}_{i}^{b}\label{eq:qt}
\end{align}
The order parameters can be understood intuitively as follows: $m_{a}$
corresponds to the overlap between the student weights $w^{a}$ and
the teacher perceptron weights. $\tilde{r}_{a}$ corresponds the norm
of expanded weights $\tilde{w}^{a}$ ; $q_{ab}$ measures the overlap
between student weight $w^{a}$ in replica and $w^{b}$ in replica
$b$. Similary $\tilde{q}_{ab}$ measures the overlap of expansion
weights $\tilde{w}^{a}$ and $\tilde{w}^{b}$ (scaled with the expansion-input
variance $\sigma_{in}^{2}$). 

We apply the replica symmetric (RS) ansatz for the order parameters
$m_{a}$, $\tilde{r}_{a}$, $q_{ab}$, and $\tilde{q}_{ab}$, which
is exact because the version space of weight vectors is connected.
This allows us write the order parameter matrices in terms of the
four scalar order parameters $m,\tilde{r},q,$ and $\tilde{q}$ as
follows
\begin{align}
m_{a} & =m,\label{eq:morder}\\
\tilde{r}_{a} & =\tilde{r},\label{eq:rtorder}\\
q_{ab} & =(1-q-\tilde{r})\delta_{ab}+q,\label{eq:qorder}\\
\tilde{q}_{ab} & =\left(\sigma_{in}^{2}\tilde{r}-\tilde{q}\right)\delta_{ab}+\tilde{q}\label{eq:qtorder}
\end{align}
 In the mean field limit, we can decompose $\avg{V^{n}}$ into the
sum of an entropic term and energetic term which are both functions
of $m,\tilde{r},q,$ and $\tilde{q}$ 
\begin{equation}
\avg{V^{n}}=\exp\left[nN(G_{0}(q,\tilde{q},\tilde{r},m)+\alpha G_{1}(q,\tilde{q},\tilde{r},m))\right]\label{eq:versionspace}
\end{equation}
Within the replica framework, the max-margin solution is the unique
solution which maximizes the margin and corresponds to the equivalence
of $w$ in all student replicas such that the overlaps $q\rightarrow1-\tilde{r}$
and $\tilde{q}\rightarrow\sigma_{in}^{2}\tilde{r}$. Taking the limit
$n\rightarrow0$, the averaged free energy takes the following functional form
\begin{eqnarray}
\langle\log V\rangle & = & N(G_{0}(\tilde{r},m)+\alpha G_{1}(\tilde{r},m))\label{eq:freenergy}
\end{eqnarray}
We obtain three closed saddle point equations for $\kappa$, $m$,
and $\tilde{r}$ by minimizing the free energy in Eqn.\ \ref{eq:freenergy}
with respect to $m$ and $\tilde{r}$ and requiring that $V\rightarrow0$.
The capacity of the network is determined by solving the mean field
equations in the limit $\kappa\rightarrow0$. 

The performance of the system depends on the expansion
parameter $\beta$, and the input and output noise variances $\sigma_{in}^{2}$
and $\sigma_{out}^{2}$. We will focus primarily on the case $\sigma_{in}=1$,
where the mean field equations simplify considerably, and solve them
for $\tilde{r}$ as a function of $m$ and $\beta$. We find that
in this case the capacity of the network as defined in terms of $\alpha_{0}$
obeys the simple scaling relation 

\begin{equation}
\alpha_{c}(\beta,\sigma_{out})=\beta\alpha_{c}(1,\sigma_{out})
\end{equation}

where $\alpha_{c}(1,\sigma_{out})$ is the capacity of the unexpanded
network shown in Fig.\ \ref{fig:capacity}. We derive an expression
for $E_{g}$ in terms of the mean field order parameters in Appendix~\ref{app:generalization}. 

\begin{figure}[h]
\includegraphics[width=0.4\textwidth]{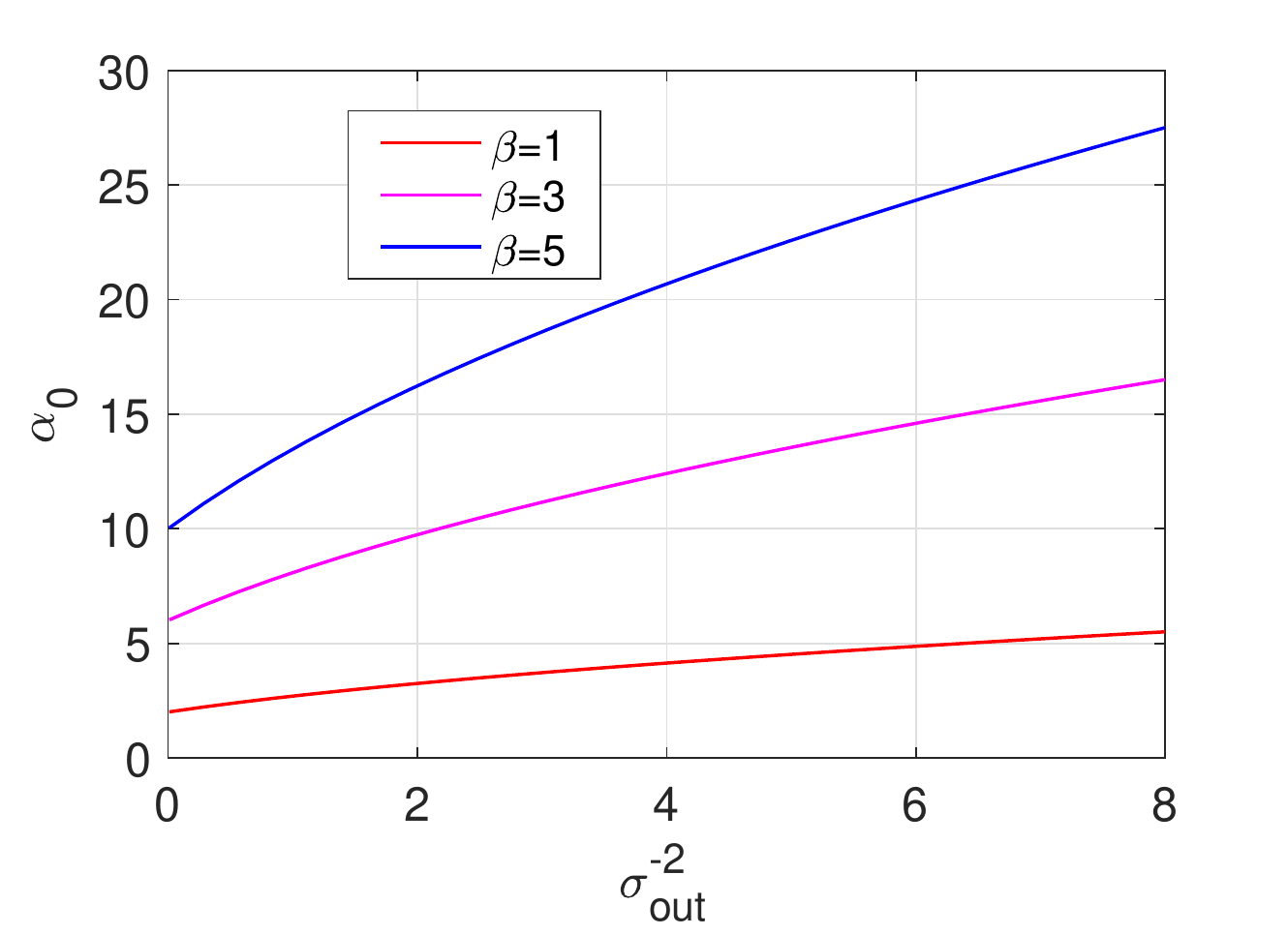}
\captionsetup{justification=raggedright,
singlelinecheck=false
}\caption{\label{fig:capacity}The network capacity for random inputs as a function
inverse variance of the label noise obtained from the solution of
the mean field equations for $\sigma_{in}=1$.}
\end{figure}

\smallskip{}
With the removal of the expanded weights $E_{g}$ takes the form:
\begin{eqnarray}
E_{g} & = & \frac{1}{\pi}\left(\frac{\pi}{2}-\tan^{-1}\left(\frac{R}{\sqrt{1+\sigma_{out}^{2}-R^{2}}}\right)\right)\label{eq:egr}
\end{eqnarray}
where $R$ is defined as the cosine of the angle between student and
teacher weights which can be expressed in terms of the order parameter
$m$ and $\tilde{r}$ as 
\begin{eqnarray}
R & = & \frac{m}{\sqrt{1-\tilde{r}}}\label{eq:rexp}
\end{eqnarray}
where the factor of $\sqrt{1-\tilde{r}}$ in Eqn.\ \ref{eq:rexp}
is the fraction of the student weight norm in the subspace of the
teacher. For student networks that are the same size as the teacher,
$\tilde{r}=0$ and $R=m$. The generalization error for a student
network that retains its expanded units after learning, with stochastic
noise included in each test example, is given by replacing $R$ with
$m$ in Eqn.\ \ref{eq:egr}. This is because the overlap $m$ is equivalent to cosine of the angle between teacher and student in the full network given the $\sqrt{N}$ normalization of both the teacher and expanded student. Thus, we see that for improved generalization
performance, it is necessary to prune the augmented units after learning
as was shown numerically for the deterministic network, Fig.\ \ref{fig:sparseprunedunpruned}.
In the stochastic expansion, the intuition for removing these weights
is straightforward as retaining them implies injecting stochastic
activities in test example, uncorrelated with the task's input, which
will obviously reduce performance. The situation is different in the
deterministic network in which correlations between the expanded and
original components of the network the network are induced by the
random map $J$, hence through learning $\tilde{w}$ acquire some
information about the task. Indeed, as we have shown above, for dense
expansion, retaining these weights slightly increases the performance.
However, for sparse expansion, the correlation between the expanded
activations and the task input is small (see below) hence pruning
improves the performance similar to the stochastic case. Finally,
we note that in the case of zero output noise, $E_{g}$ is just the
angle between the student and teacher normalized by $\pi$ and the
minimal error is given by Eqn.\ \ref{eq:egr} with $R=1$, in agreement
with Eqn.\ \ref{eq:oracle}. 

\begin{figure}[h]
	\includegraphics[width=0.5\textwidth]{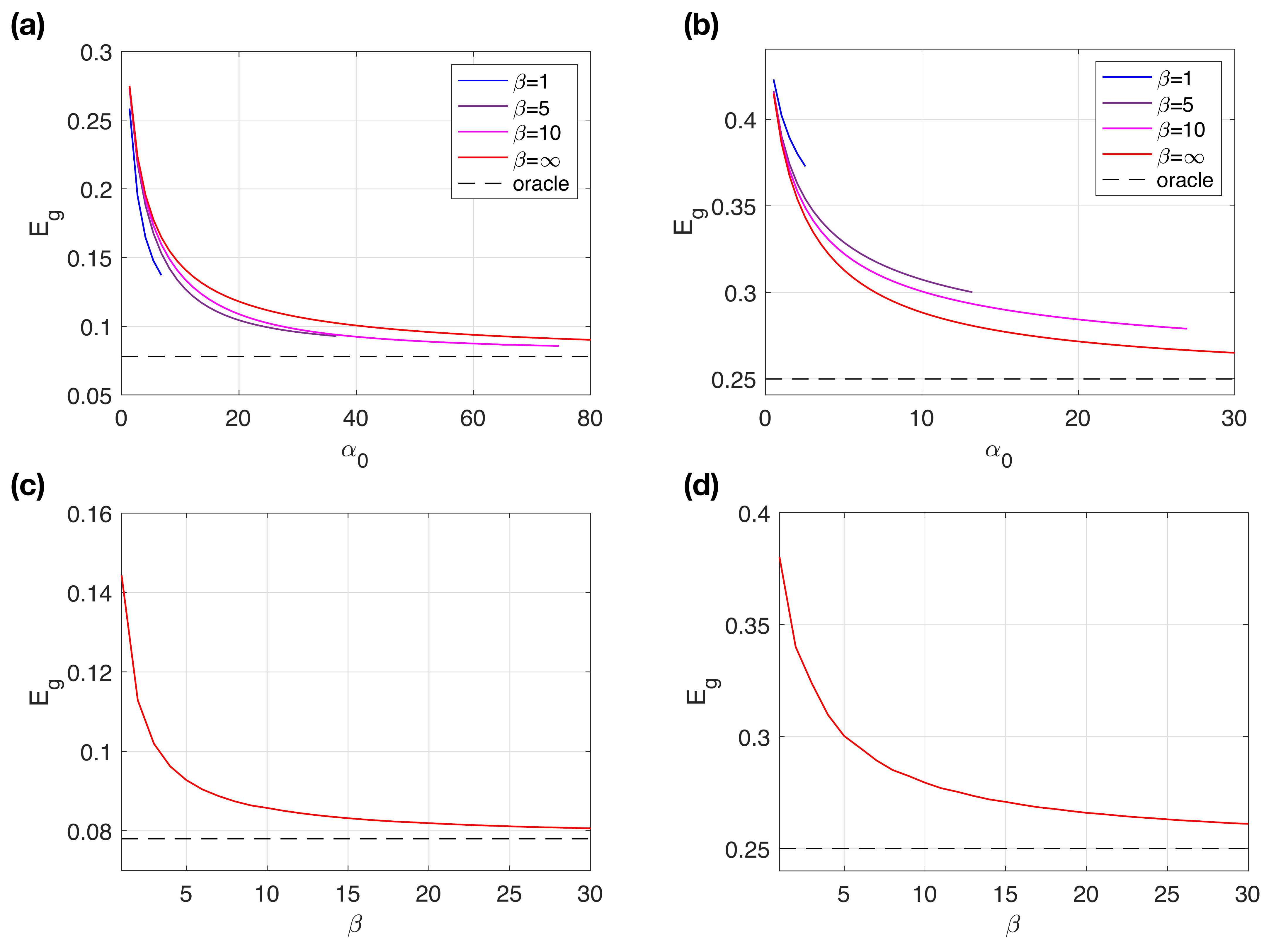}
	\captionsetup{justification=raggedright,
		singlelinecheck=false
	}\caption{\label{fig:thya} The replica theory results for the generalization
	error. $E_{g}$ is shown as a function of $\alpha_{0}$ for several
	values of the expansion factor $\beta$ for label noise with standard
	deviation $\sigma_{out}=0.25$ in (a) and standard deviation
	$\sigma_{out}=1$ in (b) . $E_{g}$ is shown as a function
	of $\beta$ for $\sigma_{out}=0.25$ in (a) and
	$\sigma_{out}=1$ in (b). $\sigma_{in}=1$ for all
	figures.}
\end{figure}

In Fig.~\ref{fig:thya}, we plot the theoretical results for $E_{g}$
as a function of $\alpha_{0}$ for different values of $\beta$ for
two values of $\sigma_{out}$ . For both high and low $\sigma_{out}$
, the generalization error decreases monotonically as a function of
$\alpha_{0}$ for fixed $\beta$ and as a function of $\beta$ for
fixed $\alpha_{0}$ . In \ref{fig:thryaabeta} and \ref{fig:thryabbeta}
of Fig.\ \ref{fig:thya} we show the minimal $E_{g}$ as a function
of $\beta$ defined as the generalization error reached for each $\beta$
after \emph{minimizing} over $\alpha_{0}$. An interesting question
is whether for a given size of training set, there is a finite optimal
expansion ratio. We find two qualitatively different behaviors dependent
on the value of $\sigma_{out}$. For low values of $\sigma_{out}$,
for a each fixed value of $\alpha_{0}$, the student network with
the lowest generalization error is the smallest network which can
fit all of the training examples. For higher values of $\sigma_{out}$,
we find that making the network larger always improves the generalization
performance for any value of $\alpha_{0}$, with the best performance
occuring in the limit $\beta\rightarrow\infty$. The crossover between
these two regimes occurs roughly around $\sigma_{out}\sim0.5$. We
conclude that adding noisy units during learning gives the network
the capacity to fit the label noise and train on more examples in
a way that does not interfere with the relevant weight information.
This allows networks with larger expansion ratios to achieve better
generalization as they are trained on more examples. 

%

\begin{figure}[h]
	\includegraphics[width=0.3\textwidth]{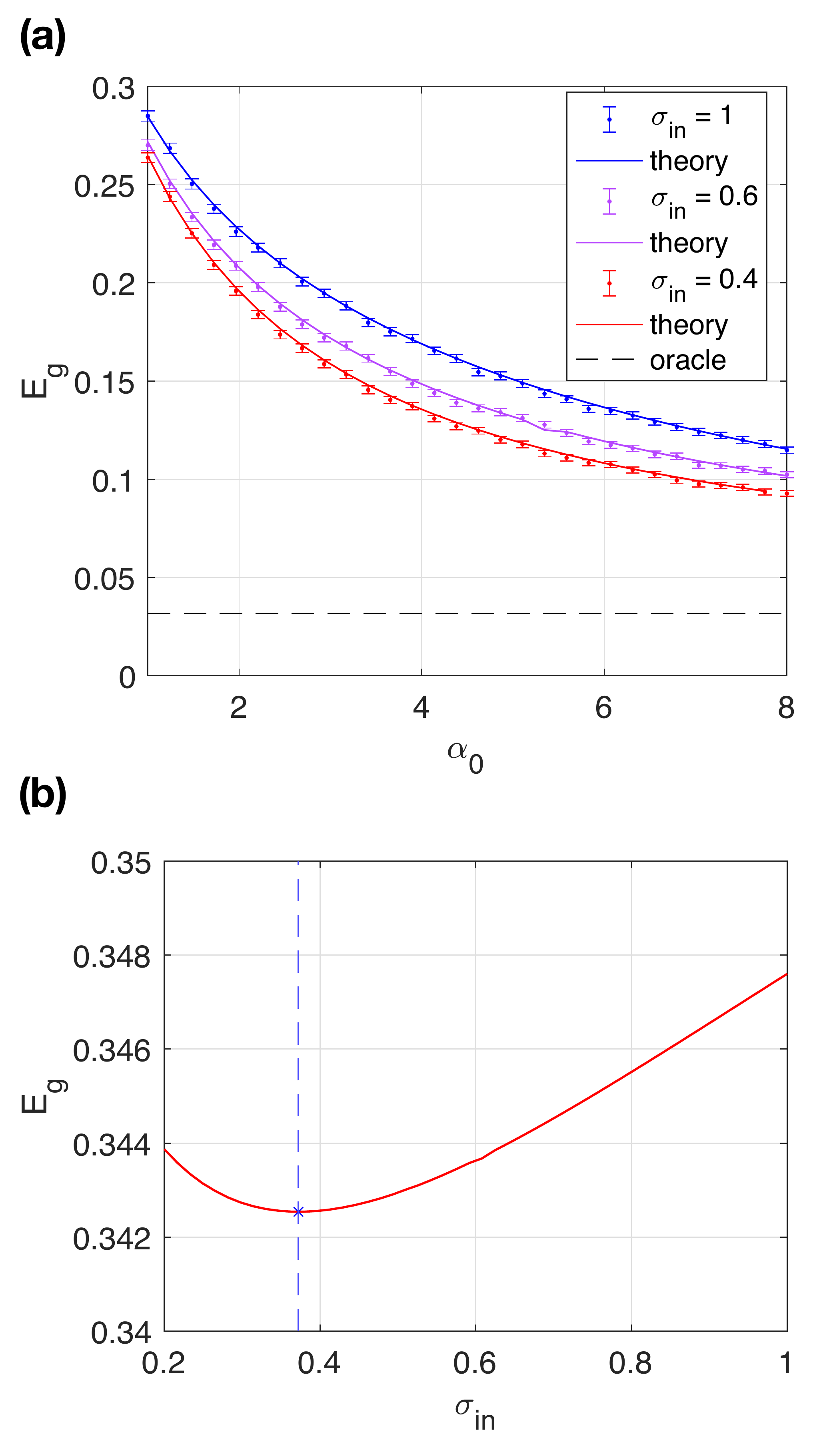}
	\captionsetup{justification=raggedright,
		singlelinecheck=false
	}\caption{\label{fig:thysim}The replica theory results compared with simulations
	for the generalization error $E_{g}$. (a) shows $E_{g}$
	for $\sigma_{out}=0.1$, $\beta=5$ and several values of $\sigma_{in}$.
	The error bars are computed from the mean and standard deviation of
	$400$ trials with $N_{0}=100.$ (b) shows $E_{g}$
	v. $\sigma_{in}$ with $\alpha_{0}=3$, $\beta=5$, and $\sigma_{out}=1$,
	$\beta=5$ and the line represents the replica predictions for the
	value of $\sigma_{in}$ that minimize the generalization error. }
\end{figure}

So far, we have considered the simple case of $\sigma_{in}=1$. We
now discuss briefly the effect of varying $\sigma_{in}$. In Fig.\ \ref{fig:thysim},
we demonstrate how varying the level of input noise can improve generalization
error by comparing theory and simulations for different choices of
$\sigma_{in}$. We find that calculations of $E_{g}$ from simulations
match very well with the value obtained from solution of the mean
field equations shown in Fig.~\ref{fig:thysim}. For low label noise,
the generalization performance is most substantially improved when
the variance of activity in the added units is much lower than the
variance of patterns being learned, i.e. $\sigma_{in}<1$.\textcolor{red}{{}
}\textcolor{black}{In the deterministic network, this corresponds
to choosing a small value for $A$. }For fixed value of label noise
$\sigma_{out}$, we find that there us an optimal variance $\sigma_{in}$
of the augmented units which minimizes $E_{g}$ for fixed measurement
density $\alpha_{0}$ and expansion factor $\beta$. This value can
be determined from the replica equations shown in Fig.~\ref{fig:thysim} (b)
and discussed in Appendix\ \ref{app:opt_input}. We will return to
this issue in the Section C.
%
\subsection{Comparison between stochastic expansion and deterministic sparse
expansion}
\label{sec:slack} 
\begin{figure}[h]
	\includegraphics[width=0.4\textwidth]{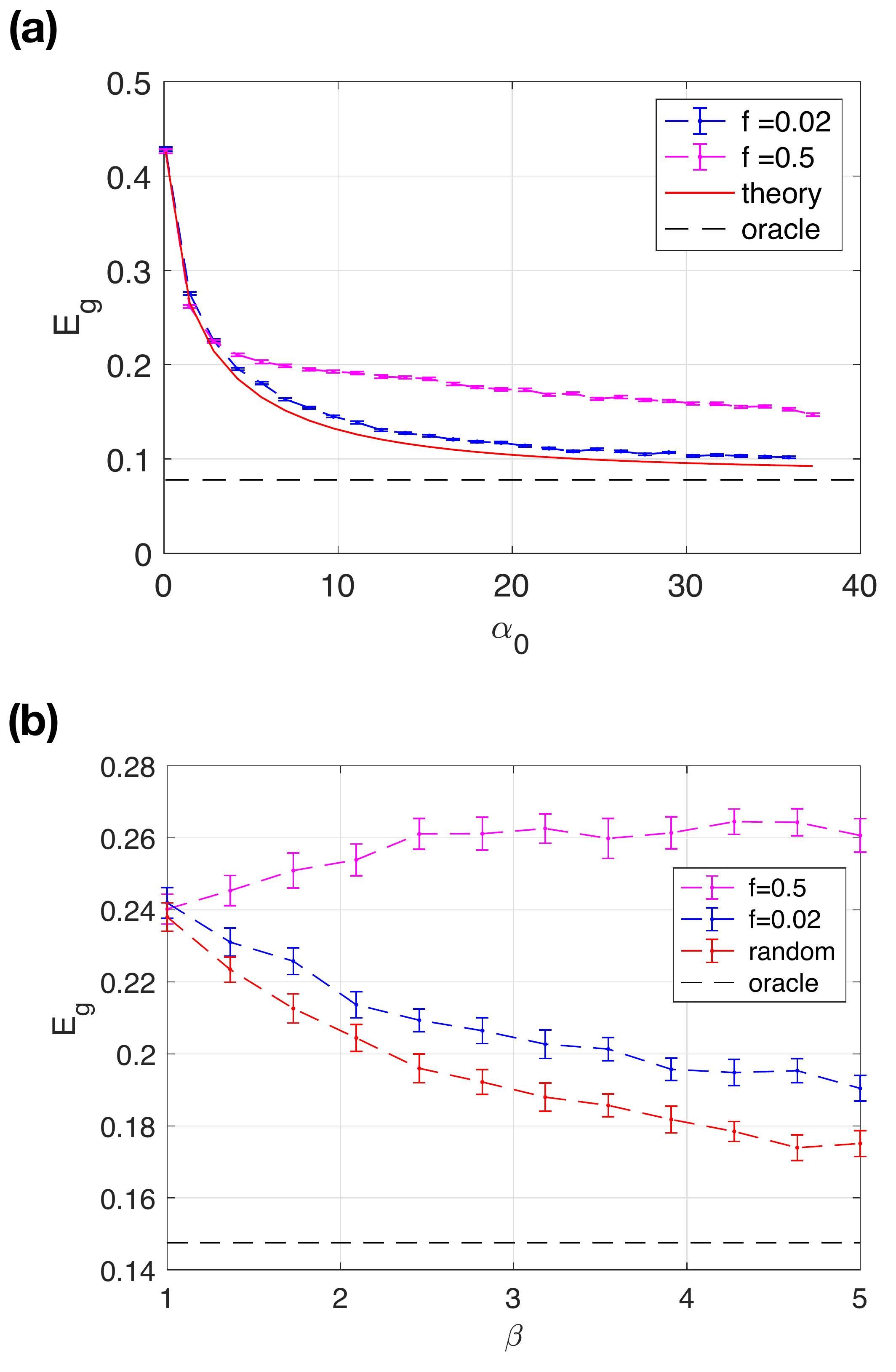}
	\captionsetup{justification=raggedright,
		singlelinecheck=false
	}\caption{\label{eq:comparisonfig} Comparison of student networks with stochastic and sparse expansions. (a) compares simulations of the two layer student
	network with the theory results for the one layer network for $\beta=5$
	for $\sigma_{out}=0.25$, $N_{0}=100$, and $200$ trials. In general,
	we see that student networks with stochastic added units attain superior
	performance when compared to a deterministic networks of the same
	size. (b) compares $E_{g}(\beta)$ for the case of stochastic
	augmented input units and deterministic hidden units with dense and
	sparse activity with $\sigma_{out}=0.5$, and $N_{0}=80$ . The parameters
	$\sigma_{in}=A=1$ and $200$ trials are used for both figures.}
\end{figure}

%


For networks expanded with sparse hidden layers, the parameter $A$
is closely related to $\sigma_{in}$. We directly compare the generalization
performance of the student network with a sparse hidden layer (Eqn.\ \ref{eq:deterministicstudent})
with the student network with stochastic units added to the input
(Eqn.\ \ref{eq:student}) by setting $\sigma_{in}=A$ so that the
statistics of the expansion units match in the two networks. For simplicity
we consider the case $\sigma_{in}=A=1$. Fig.\ \ref{eq:comparisonfig} (a) shows
the generalization error for each network with $\beta=5$ and Fig.\ \ref{eq:comparisonfig} (b)
shows the the generalization error as a function of the network expansion
factor $\beta$. The stochastically expanded network achieves superior
generalization performances for larger values of $\alpha_{0}$ and
has a higher capacity. However, as can be seen, the performance of
the deterministic networks approach that of the stochastic network
upon increasing sparsity of the hidden layer activity. This is expected
as the correlation in the sparse activities are weak and hence approach
the uncorrelated stochastic limit. 

\subsection{Correspondence with slack regularization}

\label{subsec:Correspondence-with-slack}While it is clear that expanding
a network increases its capacity, it is not obvious that the expansion
we have implemented should lead to improved generalization. While
widening a network increases its capacity to fit more training data,
it may also increase its Rademacher complexity improving its ability
to learn random input output data \cite{foundationsofmachinelearning}. However,
it turns out that the improved generalization performance in the networks
we have studied can be related to an equivalence between our expanded
network trained in the realizable regime and an unexpanded network
trained in the unrealizable regime using slack regularization \cite{10.5555/1005332.1044706,cristianini_shawe-taylor_2000}, which
we now explain.

We consider the relation between our expansion schemes for learning
and that of \emph{slack SVM }which is defined as,
\begin{equation}
\min_{w,\xi}\sum_{i=1}^{N_{0}}w_{i}^{2}+C\sum_{\mu=1}^{P}\xi^{\mu2}\enspace s.t.\enspace\enspace y_{0}^{\mu}\left(\sum_{i=1}^{N_{0}}w_{i}x_{i}^{\mu}\right)\geq1-\xi^{\mu}\label{eq:slack-1}
\end{equation}

While the SVM learning works only in the realizable regime, slack
SVM is a convex optimization that allows non zero classification errors
(when $\xi^{\mu}>1$) and regularizes them through the slack parameter
$C$ that applies $L_{2}$ regularization of the slack variables $\xi^{\mu}$
. Although it does not minimize the training error, and its cost function
does not have a well defined interpretation in terms of the classification
tasks, it is a popular learning algorithm due to its simplicity and
its empirically nice generalization properties . 

To see the relation between slack parameters and the SVM with the stochastic expansion, we first
note that the minimal $\tilde{w}$ of Eqn.\ \ref{eq:maxclass1} will
necessarily be in the span of the $P$ input stochastic vectors, $\tilde{X}_{\mu}=\widetilde{x}^{\mu}y_{0}^{\mu}$
, since any projection on the null space will increase the norm of
$\tilde{w}$ without contributing to the satisfaction of the inequalities.
Defining new variable $\xi^{\mu}$ as

\begin{eqnarray}
\xi^{\mu} & = & \tilde{X}^{\mu T}\tilde{w}
\end{eqnarray}

we can write the optimal $\tilde{w}$ as 

\begin{equation}
\tilde{w}=(\tilde{X}^{T})^{+}\xi\label{eq:wtildeconstraint-1}
\end{equation}
where $\tilde{X}$is the matrix of input stochastic vectors and $+$
denotes the pseudo-inverse operation. Substituting Eqn.\ \ref{eq:wtildeconstraint-1} into
Eqn.\ \ref{eq:maxclass1} yields

\begin{equation}
\min_{w,\xi}\sum_{i=1}^{N_{0}}w_{i}^{2}+\sum_{\mu=1}^{P}\sum_{\nu=1}^{P}\xi^{\mu}C_{\mu\nu}\xi^{\nu}\enspace\enspace s.t.\enspace\enspace y_{0}^{\mu}\left(\sum_{i=1}^{N_{0}}w_{i}x_{i}^{\mu}\right)\geq1-\xi^{\mu}\label{eq:mahalonobis}
\end{equation}

where $C=\tilde{(X}^{T}\tilde{X})^{+}$ or equivalently, 

\begin{equation}
C_{\mu\nu}=\tilde{x}^{\mu}\tilde{x}^{\nu T}\label{eq:cmatrix}
\end{equation}
which is just the sample covariance matrix of the expanded inputs
in the training set. We recognize the second term in Eqn.\ \ref{eq:mahalonobis}
as the square of the Mahalanobis distance between the vector $\xi^{\mu}$
and a set of observations with zero mean and covariance matrix $C_{\mu\nu}$
. Thus, SVM with expanded networks is equivalent to slack SVM of the
original network with a slack SVM that incorporates a Mahalanobis
distance regularization of the slack variables with a covariance regularizer
matrix $C$ injected by the expanded activities. 

Furthermore, we can establish exact correspondence between the stochastic
expansion and the slack SVM, Eqn.\ \ref{eq:mahalonobis} in the limit
large $\beta$ (and fixed $\alpha_{0}$) by noting that in this limit,
$\widetilde{x}^{\mu}\widetilde{x}^{\nu T}\sim\delta^{\mu\nu}$, hence,
the slack term becomes

\begin{equation}
\sum_{\mu=1}^{P}\sum_{\nu=1}^{P}\xi^{\mu}\langle C_{\mu\nu}^{-1}\rangle\xi^{\nu}\rightarrow\frac{1}{\sigma_{in}^{2}}\sum_{\mu=1}^{P}(\xi^{\mu})^{2}\label{eq:slack}
\end{equation}

which is a generic slack regularization term, with slack tradeoff parameter $\sigma_{in}^{-2}$.
This implies that the addition of stochastic units becomes equivalent
to the addition of slack terms in the limit $\beta\rightarrow\infty$. The equivalence breaks down for $N_{+}<P$ when the matrix
$C_{\mu\nu}$ becomes uninvertible. 

The above equivalence hold also for deterministic expansion, where
now $C_{\mu\nu}=z^{\mu}z^{\nu T}$, see Eqn.\ \ref{eq:twolayerhidden}.
In the case of a sparse expansion, $C_{\mu\nu}$ has small off-diagonal
elements and diagonal elements equaling $A$ which plays the role
of the slack regularizer.
\section{Extensions }
\label{sec:implications} So far, we have focused on a perceptron
learning a noisy perceptron rule using convex learning algorithms.
In the following section, we investigate whether random expansion
of the network during learning is beneficial when the teacher
is a given by nonlinear classification rule, and also in training
with gradient based methods. 
\subsection{Learning a nonlinear classification rule}
\label{sec:mismatch} 
\begin{figure}[h]
	\includegraphics[width=0.35\textwidth]{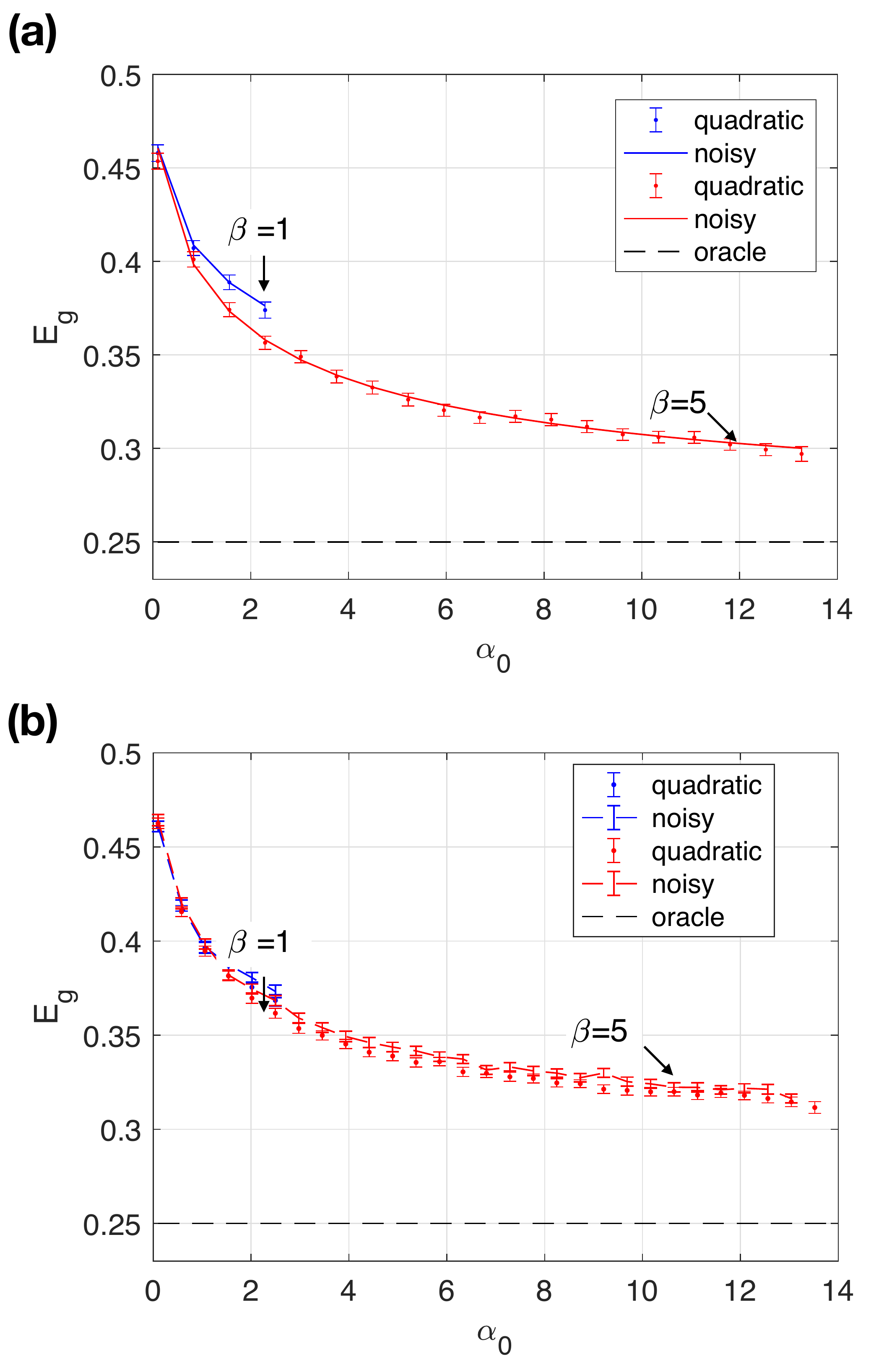}
	\captionsetup{justification=raggedright,
		singlelinecheck=false
	}\caption{\label{fig:quadteacher}Comparison of $E_{g}$ as a function of $\alpha_{0}$
	for a student learning a quadratic teacher v. the same student learning
	a linear teacher with label noise. Error bars in (a)
	are obtained from simulations of a stochastic expanded student with
	$\sigma_{in}=1$ learning a quadratic teacher for $200$ trials and
	the solid lines correspond to the replica theory result for student
	learning from a noisy teacher. In (b) we compare
	simulations of a two layer student network with $f=0.02$ and $A=1$
	learning from a quadratic teacher with simulations of the same student
	network learning from a noisy teacher for $400$ trials. The parameters
	$a=0.5$, $\sigma_{out}=1$, and $N_{0}=100$ are used in both figures.}
\end{figure}

%
%
\begin{figure}[h]
	\includegraphics[width=0.35\textwidth]{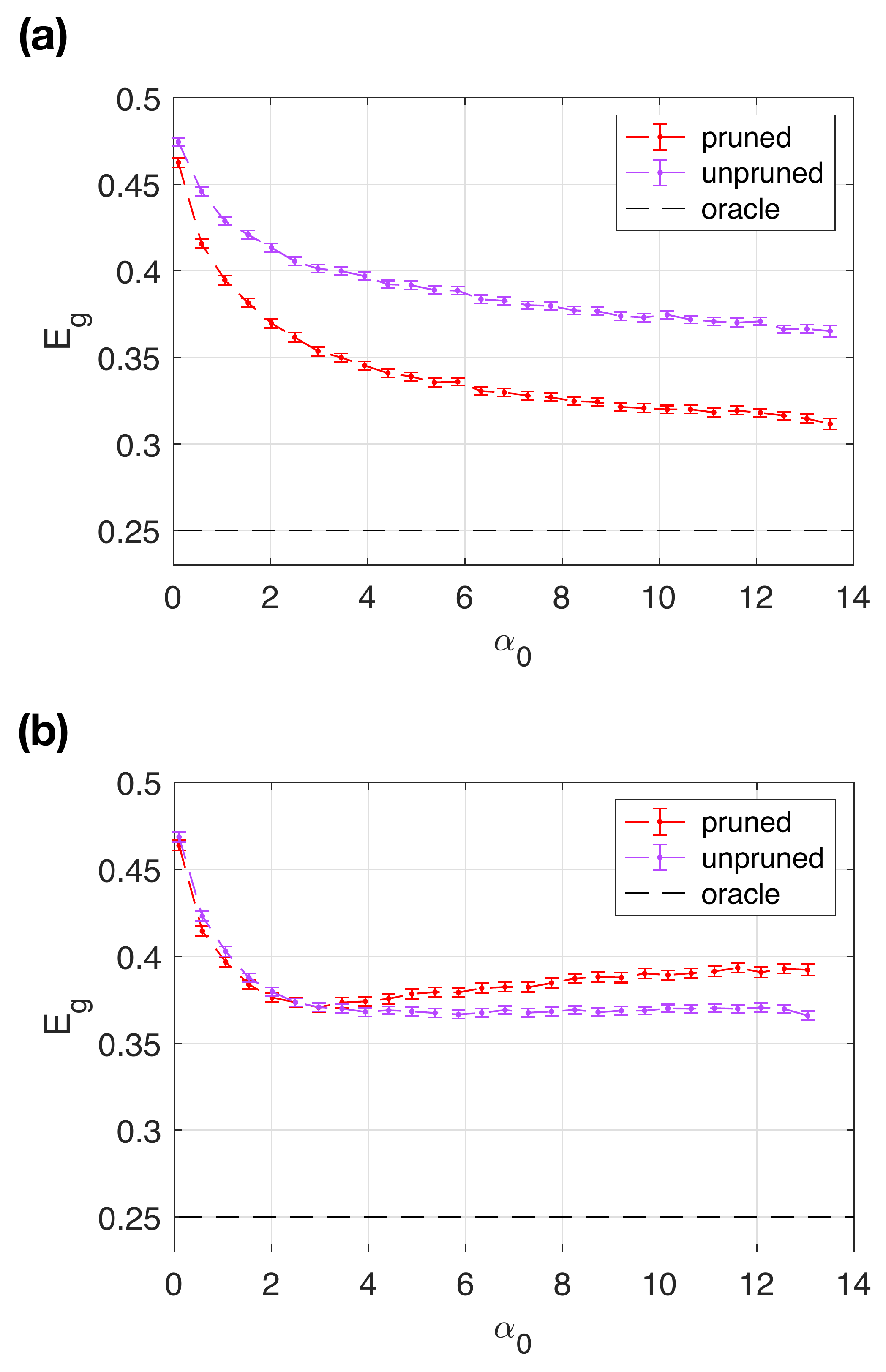}
	\captionsetup{justification=raggedright,
		singlelinecheck=false
	}\caption{\label{fig:quadsparse}Simulation results for a two layer expanded
	student learning a quadratic teacher. (a) compares
	$E_{g}$ as a function of $\alpha_{0}$ before and after pruning for
	a sparse hidden layer with $f=0.02$ and (b) compares the performance
	before and after pruning for a dense hidden layer with $f=0.5$. The parameters
	$a=0.5$, $\sigma_{out}=1$, $\beta=10$, $N_{0}=100$ and $400$
	trials are used in both figures}
\end{figure}

%

We model a perceptron learning a complex but deterministic rule by
considering a student perceptron learning from a quadratic teacher.
The target rule is then given by
\begin{equation}
y(x)=\text{sign}\left[a\frac{1}{\sqrt{N_{0}}}\sum_{i=1}^{N_{0}}{w_{i}^{0}x_{i}}+(1-a)\frac{1}{N_{0}}\sum_{i,j=1}{w_{ij}^{0}x_{i}x_{j}}\right]\label{eq:quadteacher}
\end{equation}
 with weights drawn iid as $w_{i}^{0},w_{ij}^{0}\sim\mathcal{N}(0,1)$.
Here $a$ is a scalar coefficient between zero and one and denotes the relative
weight of the linear component of the teacher. Clearly, a perceptron
student cannot emulate perfectly such a rule. For a perceptron
with $N_{0}$ weights, the optimal weights are $w=w^{0}$ with a non-zero minimal generalization error, $E_{\text{min}}$ which decreases
with $a$ . In addition, there is a critical capacity, $\alpha_{c}$
above which the training examples are unrealizable, where $\alpha_{c}$
increases with $a$ . 

We now discuss the effect of adding the stochastic
random layer as in Fig. 1D of size $N^{+}$ with $N_{0}+N^{+}=\beta N_{0}$.
Clearly the capacity for learning with zero training error increases
with $\beta$. We now ask whether this expansion is also beneficial
for generalization and whether prunning the network after learning
improves performance. We have simulated training in this network using
as before, the max-margin algorithm. Results shown in Fig.\ \ref{fig:quadteacher},
confirm that the expanded stochastic network performs better than the
unexpanded one. Furthermore, the results are in excellent agreement
with the behavior in the case of the noisy perceptron target rule,
with noise variance given by
\begin{equation}
\sigma_{out}=\frac{1-a}{a}.
\end{equation}

We show simulation results for the two layer network with dense
and sparse deterministic expansions in Fig.\ \ref{fig:quadsparse}.
\ As in the case of a noisy teacher, the optimal generalization performance
occurs after the extra neurons and synapses are removed from the network
for a sparse expansion. This effect persists for values of $\beta$
as large as $\beta=40$ for $N_{0}=60$. In the case of the two layer
network, it is not entirely obvious that removing the extra synapses
would improve performance, as this structure may be used to learn
something about the quadratic part of the teacher. It is possible
that there may be parameter regimes in which it is beneficial to keep
the extra weights unpruned that we have been unable to reach due to
computational limitations on $\beta$ and $N_{0}$. Despite these
potential shortcomings, our findings for both student architectures
demonstrate that the benefits of expanding a network can also occur
in the setting where the rule being learned is more complicated than
the model. 

\begin{figure}[h]
	\includegraphics[width=0.35\textwidth]{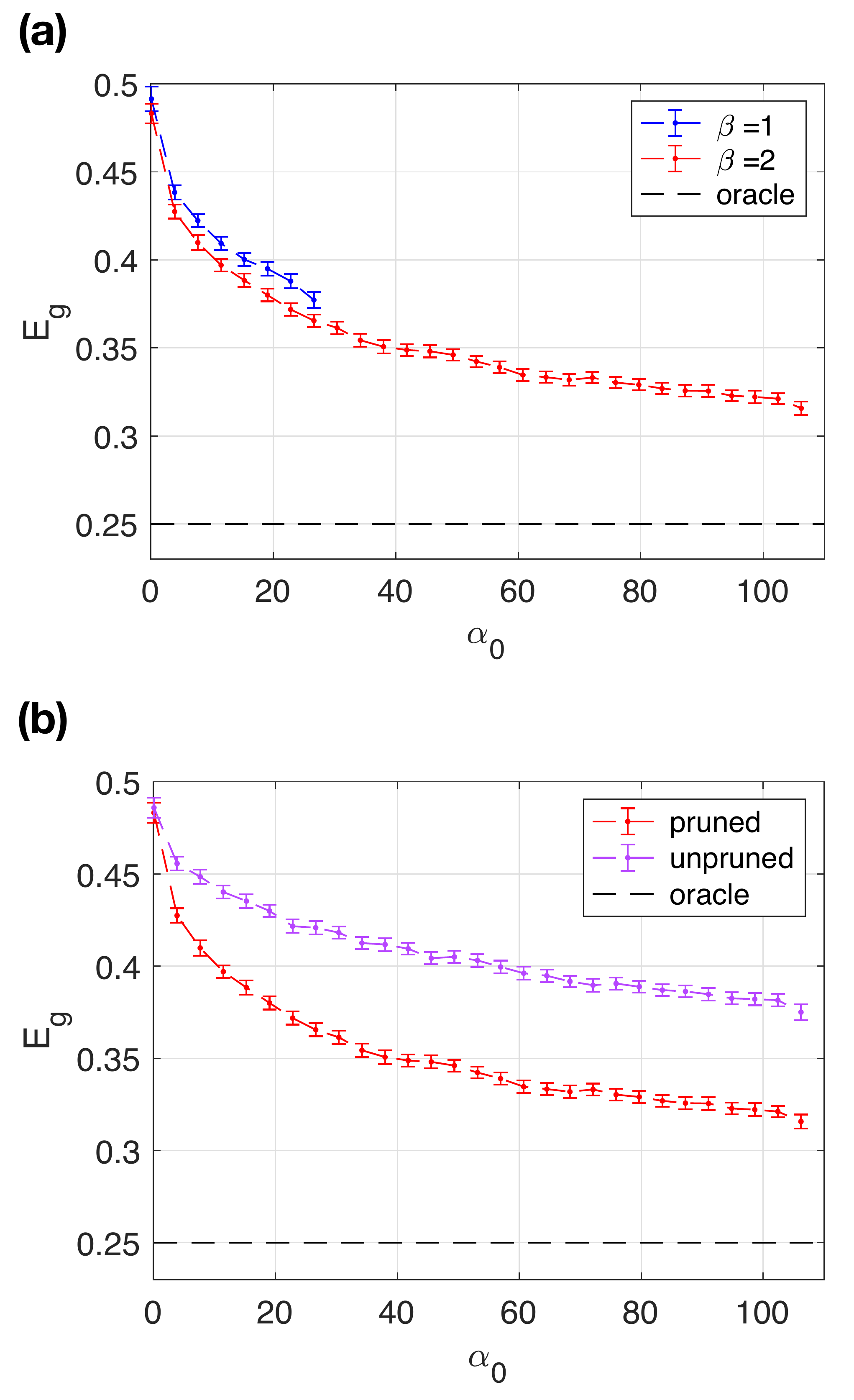}
	\captionsetup{justification=raggedright,
		singlelinecheck=false
	}\caption{\label{fig:kernel}Simulation results for a quadratic kernel student with stochastic expanded input layer
	learning a quadratic kernel teacher with label noise. The parameters $\sigma_{out}=1$, $N_{0}=20$ and $200$ trials are used in both figures.}
\end{figure}

Finally, we suggest that our results should hold in general for a nonlinear SVM teachers with field $h_{0}$ taking the following form
\begin{align}
	h_{0}=w^{0}\cdot\Phi(x)+\epsilon
\end{align}
where $\Phi(x)$ is a transformation from $\mathbb{R}^{N_{0}}\rightarrow \mathbb{R}^{M_{0}}$ defined by $\Phi(x)=(\Phi_{1}(x),\dots,\Phi_{M_{0}}(x))$. 
Given a training set $(x^{\mu},y^{\mu}_{0})$, we can consider a student with labels given by
\begin{align}
y=\text{sign}(w\cdot \Psi(x))	
\end{align}
where $\Psi(x)$ is a transformation from $\mathbb{R}^{N}\rightarrow \mathbb{R}^{M}$ defined by $\Psi(x)=(\Psi_{1}(x),\dots,\Psi_{M}(x))$. The max margin solution for the student weight vector is then
\begin{align}
w=\sum_{\mu=1}^{P}\alpha_{\mu}y^{\mu}_{0}\Psi(x^{\mu}+\tilde{x}^{\mu})	
\end{align}
where the coefficients $\alpha^{\mu}$ are given by solving the optimization problem \ref{eq:maxclass1} with constraints given by \ref{eq:maxmarginclassifier}. The student labels are now given by
\begin{align}
y&=\text{sign}\left(\sum_{\mu=1}^{P}\alpha_{\mu}y_{0}^{\mu}K(x^{\mu}+x,x+\tilde{x})\right)
\end{align}
where the kernel $K$ is defined as 
\begin{align}
	K(x,y)=\Psi(x)\cdot \Psi(y)
\end{align}
If the student perceptron uses the same transformation $\Phi(x)$ as the teacher, expanding the dimensionality of the input $x$ produces the same improvement in generalization performance as shown in section~\ref{sec:perceptrontheory}. This is because the training data is linearly separable without noise. We demonstrate this with simulations with $\Phi(x)$ chosen as the eigenfunction of a quadratic kernel with a stochastic expansion of the input shown in Fig.\ref{fig:kernel}. When the transformation $\Psi(x)$ is less complex than the transformation $\Phi(x)$ of the teacher, we expect that expanding the input of the student results in improved generalization qualitatively similar to our results for a linear student learning a quadratic teacher discussed in the beginning of the section. We also expect these results to hold for networks with sparse expansions.

%
\subsection{Logistic regression}
\label{sec:otheropt} 
We now consider alternative optimization methods and loss functions
which allow a neural network to be trained beyond capacity. One example
is logistic regression, with a cost function given by 
\begin{eqnarray}
L(w) & = & \sum_{\mu=1}^{P}\log\left(1+\exp(-u^{\mu})\right)\\
u^{\mu} & = & \frac{1}{\sqrt{N}}\sum_{i=1}^{N}y^{\mu}w_{i}x_{i}^{\mu}
\end{eqnarray}

In the following, we consider full batch gradient descent so that
the update to the weights at each training epoch is given by 

\begin{align*}
\Delta w_{i}= & -\eta\frac{\partial L(w)}{\partial w_{i}}
\end{align*}

In \cite{Soudry2018} it was shown that the normalized weight vector
obtained by a minimizing the logistic regression loss function via
gradient descent should converge to the max margin solution after
a sufficiently long training time if the training data is linearly
separable. However, this correspondence depends on learning parameters
such as $\eta$ and the number of iterations. Note that in general,
for convergence to the max margin solution one needs to run the logistic
regression gradient based training for longer times than required for
finding a solution with zero training error. For unrealizeable rules,
e.g., the noisy teacher in Eqn.\ \ref{eq:trueteacher} and the quadratic
teacher Eqn.\ \ref{eq:quadteacher}, logistic regression and max-margin
classification are not equivalent for large $P$ because the training
set provided by the teacher is not linearly separable. 

In previous sections we have shown that stochastic and sparse expansions
of perceptron networks increase the capacity of a network by making
the training set linearly separable in a higher dimensional space.
Thus, it is natural to ask under what conditions training an expanded
network via logistic regression will result in a weight vector that
converges to the new max margin solution in the higher dimensional
space and if this solution can yield a superior generalization performance
compared to a gradient based training of the unexpanded student network. 

\begin{figure}
\includegraphics[width=0.35\textwidth]{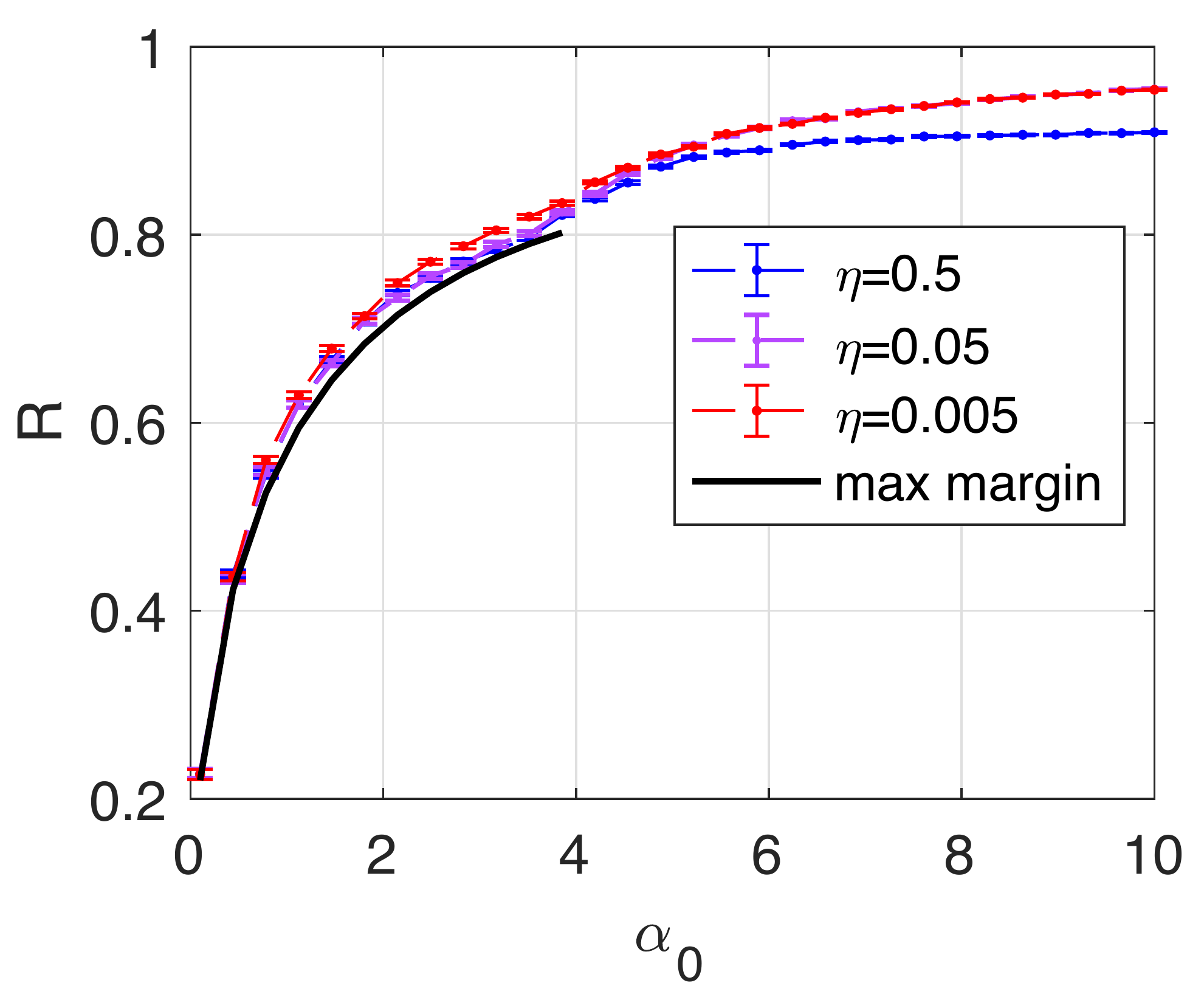} 
\captionsetup{justification=raggedright,
singlelinecheck=false
}
\caption{\label{fig:Simulation-results-for}Simulation results for logistic
regression showing $R$ v. $\alpha_{0}$ for various learning rates
for $N=100$, and $\sigma_{out}=0.5$. }
\end{figure}

\begin{figure}[h]
	\includegraphics[width=0.45\textwidth]{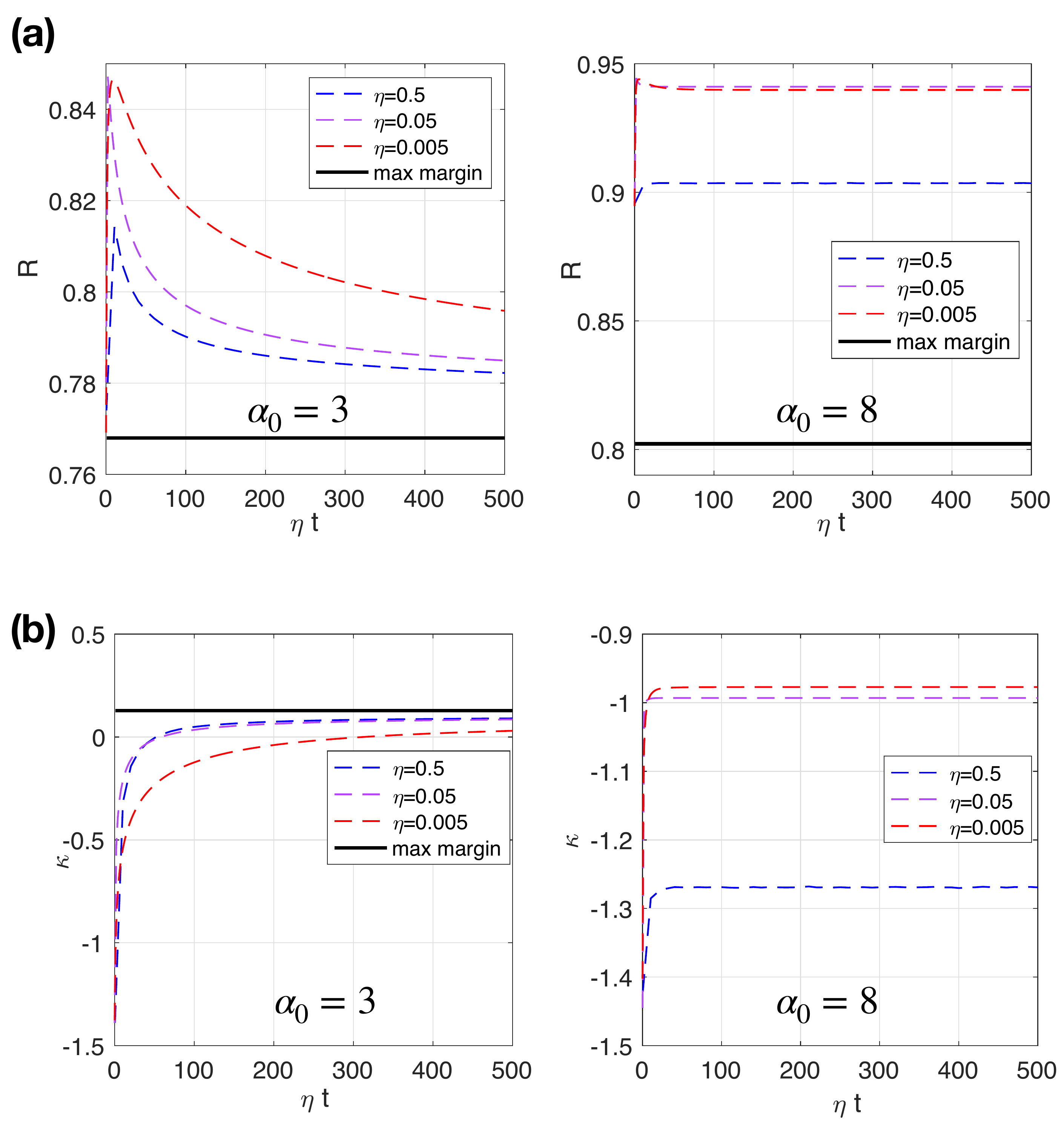}
	\captionsetup{justification=raggedright,
		singlelinecheck=false
	}\caption{\label{fig:logisticbeta1}Simulation results for logistic regression
	with $\beta=1$ and $\sigma_{out}=0.5$ where $t$ is defined as the
	number of training epochs. (a) shows $R$ as a
	function of $\eta t$ where for $\alpha_{0}=3$ (below capacity) and
	$\alpha_{0}=8$ (above capacity) for $N=100$.(b)
	shows the margin $\kappa$ as a function of $\eta t$ for $\alpha_{0}=3$
	and $\alpha_{0}=8$.}
\end{figure}

We have simulated the logistic regression learning for the problem
of learning a noisy perceptron teacher, for some values of $\eta$
and number of training epochs. We first consider the case $\beta=1$,
i.e. a student the same size as the teacher. For $\alpha_{0}$ below
capacity, the margin increases monotonically with training epochs
and converges asymptotically to the maximum margin, as shown in Fig.\ \ref{fig:logisticbeta1} (b)
($\alpha_{0}=3$), with convergence time depending on $\eta$. In
Fig.\ \ref{fig:logisticbeta1} (a) we show for the same $\alpha_{0}$
the value of the overlap between student and teacher, as a function
of $\eta t$. Interestingly, while $R$ does seem to converge asymptotically
to the maximum margin value, it is not monotonic and in fact reaches
a maximum value larger than the infinite time asymptote early in
the training. Thus, the max margin solution is not necessarily the
one with the best generalization performance. Above capacity, logistic
regression permits solutions with nonzero training error, and we find
that it results with good generalization performance. The value of
$R$ as a function of $\alpha_{0}$ is shown in Fig.\ \ref{fig:Simulation-results-for}.
As seen, for small $\alpha_{0}$ the overlap (achieved after a large
number of epochs) is close to the max margin solution with precise
values dependent on $\eta$ and the stopping criterion. When $\alpha_{0}$
increases above capacity, $R$ increases monotonically and seems to
approach $R=1$ for large $\alpha_{0}$ (corresponding to the optimal
solution $w=w^{0}$), although the amount of increase depends on $\eta$.
Note that in this regime both $R$ and $\kappa$ converge fast to
their asymptotic values as shown in Fig.\ \ref{fig:logisticbeta1}. 

%
%
%

\begin{figure}
\includegraphics[width=0.4\textwidth]{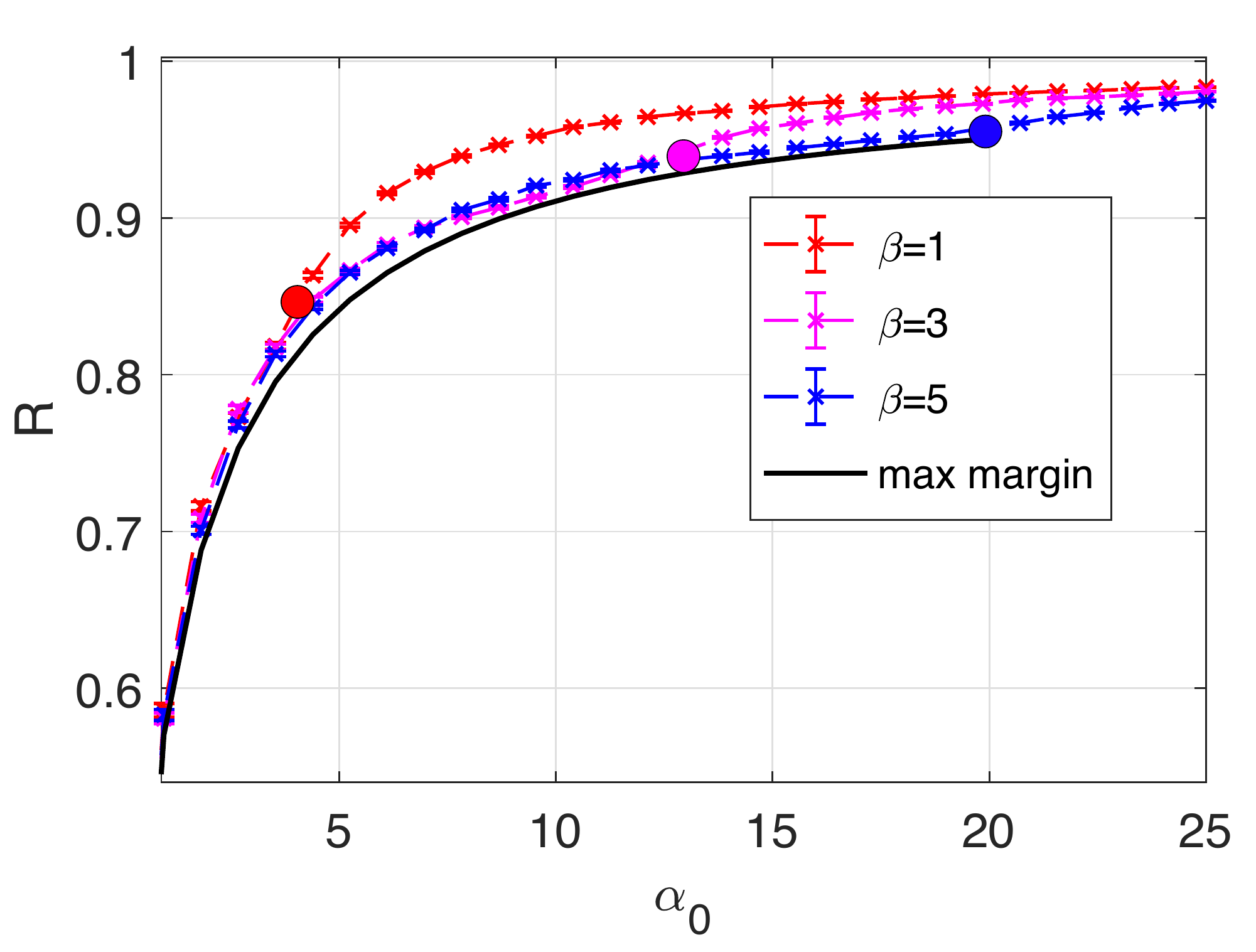}\captionsetup{justification=raggedright,
singlelinecheck=false
}\caption{\label{fig:logisticbeta135}Simulation results for logistic regression
for several values of $\beta$ with $N_{0}=100$ and $\sigma_{out}=0.5$
showing $R$ v. $\alpha_{0}$ for $\beta=1,3,5$ with fixed learning
rate $\eta=0.01$. The max margin line in the plot corresponds to
the max margin solution for $\beta=5$ and the circles mark the capacity
for each value of $\beta$.}
\end{figure}

For an expanded student network i.e. $\beta>1,$ we find that $R$
converges to the max margin value after long training time for $\alpha_{0}$
below capacity as shown in Fig.\ \ref{fig:logisticbeta135} and continues
to increase with $\alpha_{0}$ as it increases above capacity. However,
for fixed values of $\eta$ that are not too large, the largest value
of $R$ for any $\alpha_{0}$ is obtained for the unexpanded network,
i.e., $\beta=1$ as shown in Fig.\ \ref{fig:logisticbeta135}. This
implies that in this range of parameters, expanding the network does
not improve generalization performance. 

\section{Discussion}

In this work, we have shown how expanding the architecture of neural
networks can provide computational benefits beyond better expressivity
and improve the generalization performance of the network after the
expanded weights and neurons are pruned after training. We obtain
equations for the order parameters characterizing generalization in
randomly expanded perceptron networks (called stochastic expansion)
in the mean limit and show explicitly that expansion allows for more
accurate learning of noisy or complex teacher networks. This is achieved
by increasing network capacity during training, allowing the learning
to benefit from more examples. We show a qualitatively similar improved
performance when expanding by adding fixed random weights (deterministic
expansion) connecting the input to sparsely active hidden units. An
additional insight into our results is provided by showing that the
expansion is effectively similar to the addition of slack variables
to the max-margin learning. 

In our analysis, we considered training sets drawn iid from a Gaussian
distribution with no spatial structure. It would be interesting to
see how our results could be extended to learning structured data.
In particular, \cite{PhysRevX.8.031003} developed a theory for the
linear classification of manifolds with arbitrary geometry by using
special anchor points on the manifolds to define novel geometrical
measures of radius and dimension which can be directly linked to the
classification capacity for manifolds of various geometries. It would
be interesting to see if sparse expansions similar to those we have
studied could be useful in classifying noisy manifolds and if there
is any correspondence to SVMs containing anisotropic slack regularization
encoded in the structure of the covariance matrix as in Eqn.\ \ref{eq:cmatrix}. 

It would also be interesting to determine how and if our observations
apply to learning in deep networks with multiple layers. Neural network
pruning techniques have been widely discussed in the deep learning
community and it has been shown that neural network pruning
techniques can reduce parameter counts of trained network by over
90\% without compromising accuracy \cite{NIPS1989_250,NIPS2015_5784}.
Training a pruned model from scratch is worse than retraining a pruned
model, which suggests that the extra capacity of the network allows
it to find more optimal solutions. In \cite{frankle2018the}, the
authors find that dense, randomly-initialized, feed-forward networks
contain subnetworks that can reach test accuracy comparable to the
original network in a similar number of training iterations when trained
in isolation. It would be interesting to see if the extra weights
in the larger networks can be translated into a regularization condition
on the subnetwork.

Most of our work focused on max margin learning. We have explored
the effect of expansion on gradient based learning with logistic regression
cost function. We find that for appropriate choice of learning rate
and learning time, generalization is similar to the max margin performance
below the network capacity, consistent with \cite{Soudry2018}. We
also found that in the explored parameter range, optimal generalization
performance is achieved by the unexpanded network, as gradient based
learning can extract useful information even beyond the capacity learning.
However, understanding the generalization performance in gradient
based learning requires a more thorough understanding of the role
of learning rate and training time is quite difficult given the lack
of theory for the training dynamics for logistic regression. It would
be interesting to see if there is a way to scale $\eta$ such that
expanding the network can provide similar benefits for logistic regression
beyond capacity as for max margin learning. We leave this to future
work. 

We also note that generalization can also improve when adding unquenched
noise to the student labels during training with logistic loss as
this prevents the classifier from overfitting (results not shown;
\cite{doi:10.1162/neco.1995.7.1.108,Welling:2011:BLV:3104482.3104568}).
This differs from our construction for two reasons. The first is that
our student by construction learns the weights in the extended
part of the network. The second is that our dimensionality expansion
changes the properties of the training set in that a nonlinearly
separable training set in the original space may become linearly separable
in the higher dimensional expanded space.
\section{Acknowledgments}
We thank Haozhe Shan and Weishun Zhong for valuable
discussions concerning our logistic regression results and Subir Sachdev
for helpful comments on the draft. This work is partially supported
by the Gatsby Charitable Foundation, the Swartz Foundation, and the National
Institutes of Health (Grant No. 1U19NS104653). J.S. acknowledges support
from the National Science Foundation Graduate Research Fellowship 
under Grant No. DGE1144152.
\bibliography{noisyperceptron.bib}
\appendix
\pagebreak\onecolumngrid\newpage{}
\section{Mean field equations}
\label{app:replica} We outline the derivation of the mean field equations
used to compute the order parameters defined in Eqns.~\ref{eq:m},\ref{eq:rt},
\ref{eq:q}, and \ref{eq:qt} which are used to compute the generalization
error given in Eqn.\ \ref{eq:egr}. We define the student field for
each replica of the student network as: 
\begin{equation}
h^{\mu a}=\frac{1}{\sqrt{N}}\sum_{i=1}^{N}W_{i}X{}^{\mu}=\frac{1}{\sqrt{N}}\left(\sum_{i=1}^{N_{0}}w_{i}^{a}x_{i}^{\mu}+\sum_{j=1}^{N_{+}}\tilde{w}_{j}^{a}\tilde{x}_{j}^{\mu}\right),\label{eq:studentfield}
\end{equation}
and the teacher field as 
\begin{equation}
h^{\mu0}=\frac{1}{\sqrt{N}}\sum_{i=1}^{N}W_{i}^{0}\cdot X_{i}{}^{\mu}+\epsilon^{\mu}=\frac{1}{\sqrt{N}}\sum_{i=1}^{N_{0}}w_{i}^{0}x_{i}^{\mu}+\epsilon^{\mu}.\label{eq:teacherfield}
\end{equation}

We can now write the average over the version space in Eqn.~\ref{eq:versionspace}
in terms of these new variables 
\begin{equation}
V^{n}=\left\langle \int\prod_{i=1}^{N_{0}}\prod_{j=1}^{N_{+}}\prod_{a}{dw_{i}^{a}d\tilde{w}_{j}^{a}\delta\left(\sum_{i=1}^{N_{0}}(w_{i}^{a})^{2}+\sum_{j=1}^{N_{+}}(\tilde{w}_{j}^{a})^{2}-N\right)}\prod_{\mu}^{P}\int dh^{\mu a}\int d\hat{h}^{\mu a}\int dh_{0}^{\mu}\int d\hat{h}_{0}^{\mu}\left[\sum_{\sigma}\Theta_{\mu,a}\left(yh^{\mu a}-\kappa\right)\Theta(yh_{0}^{\mu})\right]I\right\rangle 
\end{equation}
where $I$ is given by 
\begin{eqnarray}
I & = & \exp\left[-i\sum_{a\mu}h^{\mu a}\hat{h}^{\mu a}-i\sum_{\mu}h_{0}^{\mu}\hat{h}_{0}^{\mu}+i\sum_{a\mu}\hat{h}^{\mu a}\frac{1}{\sqrt{N}}\sum_{i=1}^{N}\left(W_{i}^{a}X_{i}^{\mu}\right)+i\sum_{\mu}\hat{h}_{0}^{\mu}\left(\frac{1}{\sqrt{N}}\sum_{i=1}^{N}W_{i}^{0}\cdot X_{i}^{\mu}+\epsilon^{\mu}\right)\right]
\end{eqnarray}
and the constraints in Eqns.~\ref{eq:studentfield} and \ref{eq:teacherfield}
are implemented by the Lagrange multipliers $h^{\mu a}$ and $\hat{h}_{0}^{\mu}$.
Averaging over the input $x^{\mu}$, $\tilde{x}$, and the noise $\epsilon^{\mu}$,
$I$ becomes 

\begin{align}
I= & \int\prod_{\mu=1}^{P}\prod_{i=1}^{N_{0}}\frac{dx_{i}^{\mu}}{\sqrt{2\pi}}e^{-\frac{(x^{\mu})^{2}}{2}}\prod_{j=1}^{N_{+}}\frac{d\tilde{x}_{j}^{\mu}}{\sqrt{2\pi\sigma_{in}^{2}}}e^{-\frac{(\tilde{x}^{\mu})^{2}}{2\sigma_{in}^{2}}}\frac{d\epsilon^{\mu}}{\sqrt{2\pi\sigma_{out}^{2}}}e^{-\frac{(\epsilon^{\mu})^{2}}{2\sigma_{out}^{2}}}\nonumber \\
\times & \exp\left[-i\sum_{\mu\alpha}h^{\mu a}\hat{h}^{\mu a}-i\sum_{\mu}h_{0}^{\mu}\hat{h}_{0}^{\mu}+i\sum_{\mu a}\sum_{i=1}^{N_{0}}\frac{1}{\sqrt{N}}(\hat{h}^{\mu a}w_{i}^{a}+\hat{h}_{0}^{\mu}w_{i}^{0})x_{i}^{\mu}+i\sum_{\mu a}\sum_{j=1}^{N_{+}}\frac{1}{\sqrt{N}}(\hat{h}^{\mu a}\tilde{w}_{j}^{a}\tilde{x_{j}}^{\mu})+i\sum_{\mu}\hat{h}_{0}^{\mu}\epsilon^{\mu}\right]\nonumber \\
= & \exp\left[-\sum_{\mu}\Big(i\sum_{a}h^{\mu a}\hat{h}^{\mu a}+ih_{0}^{\mu}\hat{h}_{0}^{\mu}+\sum_{a}\hat{h}^{\mu a}\hat{h}_{0}^{\mu}\sum_{i=1}^{N_{0}}\frac{w_{i}^{a}\cdot w_{i}^{0}}{N}+\frac{(1+\sigma_{out}^{2})}{2}\hat{h}_{0}^{\mu2}+\frac{1}{2}\sum_{ab}\hat{h}^{\mu a}\hat{h}^{\mu b}\left(\sum_{i=1}^{N_{0}}\frac{w_{i}^{a}w_{i}^{b}}{N}+\sigma_{in}^{2}\sum_{j=1}^{N_{+}}\frac{\tilde{w}_{j}^{a}\tilde{w}_{j}^{b}}{N}\right)\Big)\right]
\end{align}

We define the order parameters $m_{a}$, $q_{ab}$ and $\tilde{q}_{ab}$
as 
\begin{align}
m_{a} & =\frac{1}{N}\sum_{i=1}^{N_{0}}w_{i}^{0}w_{i}^{a}\\
q_{ab} & =\frac{1}{N}\sum_{i=1}^{N_{0}}w_{i}^{a}w_{i}^{b}\\
\tilde{q}_{ab} & =\frac{\sigma_{in}^{2}}{N}\sum_{j=1}^{N_{+}}\tilde{w}_{j}^{a}\tilde{w_{j}}^{b}
\end{align}
 For further convenience, we write the sum of $q_{ab}$ and $\tilde{q}_{ab}$
as 
\begin{equation}
Q_{ab}=q_{ab}+\tilde{q}_{ab}.
\end{equation}
In terms of the order parameters, $I$ becomes 
\begin{eqnarray}
I & = & \exp\left[-\sum_{\mu}\Big(i\sum_{\alpha}h^{\mu a}\hat{h}^{\mu a}+ih^{\mu0}\hat{h}^{\mu0}+\sum_{a}\hat{h}^{\mu a}\hat{h}^{\mu0}m_{a}+\frac{(1+\sigma_{out}^{2})}{2}(\hat{h}^{\mu0})^{2}+\frac{1}{2}\sum_{ab}\hat{h}^{\mu a}\hat{h}^{\mu b}Q_{ab}\Big)\right]
\end{eqnarray}
We can now do the integrals over $\hat{h}$ and $\hat{h}_{0}$ which
gives us
\begin{eqnarray}
\prod_{\mu}^{P}\int dh^{\mu a}\int d\hat{h}^{\mu a}\int dh^{\mu0}\int d\hat{h}^{\mu0}I & = & \prod_{\mu}^{P}\int dh^{\mu a}\int D\bar{h}^{\mu0}\det(Q_{ab}-\bar{m}^{2})^{-\frac{P}{2}}X
\end{eqnarray}
where we have defined $X$ as 
\begin{eqnarray}
X & = & \exp\left[-\frac{1}{2}\sum_{\mu}\sum_{ab}(\bar{h}^{\mu0}\bar{m}-h^{\mu a})(Q_{ab}-\bar{m}^{2})^{-1}(\bar{h}^{\mu0}\bar{m}-h^{\mu b})\right]\label{eq:qmatrix}
\end{eqnarray}
and $\bar{m}$ and $\bar{h}$ as 
\begin{align}
\bar{m} & =\frac{m}{\sqrt{1+\sigma_{out}^{2}}}\\
\bar{h}^{0} & =\frac{h^{0}}{\sqrt{1+\sigma_{out}^{2}}}
\end{align}
We now define the additional parameter $\tilde{r}_{a}$ as 
\begin{align}
\tilde{r}_{a} & =\frac{1}{N}\sum_{i=1}^{N_{+}}(\tilde{w}_{i}^{a})^{2}
\end{align}
Since the solution space is connected, we can make the following replica
symmetric ansatz for $m_{a}$, $q_{ab}$, $\tilde{q}_{ab}$, and $\tilde{r}_{a}$
\begin{align}
m_{a} & =m\label{eq:replicasym}\\
\tilde{r}_{a} & =\tilde{r}\label{eq:replicasym2}\\
q_{ab} & =(1-\tilde{r}-q)\delta_{ab}+q\label{eq:replicasym3}\\
\tilde{q}_{ab} & =\left(\sigma_{in}^{2}\tilde{r}-\tilde{q}\right)\delta_{ab}+\tilde{q}\label{eq:replicasym4}\\
Q_{ab} & =(r_{Q}-Q)\delta_{ab}+Q
\end{align}
where $Q=q+\tilde{q}$ and $r_{Q}=1-(1-\sigma_{in}^{2})\tilde{r}$.
The inverse of the matrix in Eqn.~\ref{eq:qmatrix} is given by
\begin{eqnarray}
(Q_{ab}-\bar{m}^{2})^{-1}=\frac{1}{r_{Q}-Q}\delta_{ab}-\frac{Q-\bar{m}^{2}}{(r_{Q}-Q)^{2}}
\end{eqnarray}

we now define $X^{\prime}$ as: 
\begin{eqnarray}
X^{\prime} & = & \prod_{\mu}^{P}\int dh^{\mu a}\int dh_{0}^{\mu}\exp\Big[-\frac{P}{2}\log\det(Q_{ab}-\bar{m}^{2})\Big]X^{P}
\end{eqnarray}
Plugging in the replica symmetric ansatz in Eqns.~\ref{eq:replicasym},
\ref{eq:replicasym2}, \ref{eq:replicasym3}, \ref{eq:replicasym4},
this becomes 

\begin{align}
X^{\prime} & =\prod_{\mu}^{P}\int dh^{\mu a}\int dh^{\mu0}\exp\Big[-\frac{1}{2(r_{Q}-Q)}\sum_{\mu a}(h^{\mu a})^{2}+\frac{1}{2}\frac{Q-\bar{m}^{2}}{(r_{Q}-Q)^{2}}\sum_{\mu}\left(\sum_{a}h^{\mu a}\right)^{2}\nonumber \\
 & +\frac{1}{(r_{Q}-Q)}\sum_{\mu}\bar{h}^{\mu0}\bar{m}\sum_{a}h^{\mu a}-\frac{n\sum_{\mu}(\bar{h}^{\mu0}\bar{m})^{2}}{2(r_{Q}-Q)}-\frac{P}{2}\log\det(Q_{ab}-\bar{m}^{2})\Big]\label{eq:xprime}
\end{align}

We decouple terms with different replica indices in Eqn.~\ref{eq:xprime} via a Hubbard-Stratonovich transformation
by introducing the auxiliary variable $t$. Then $X^{\prime}$ becomes
\begin{align}
X^{\prime} & =2\int_{0}^{\infty}D\bar{h}_{0}\int Dt\left[\int_{\kappa}^{\infty}\frac{dh}{\sqrt{2\pi}}\exp\Big(-\frac{1}{2}\frac{h^{2}}{r_{Q}-Q}+\frac{\sqrt{Q-\bar{m}^{2}}}{r_{Q}-Q}ht+\frac{1}{r_{Q}-Q}h\bar{h}_{0}\bar{m}-\frac{\bar{m}^{2}\bar{h}_{0}^{2}}{2(r_{Q}-Q)}\Big)\right]^{n}
\end{align}
where $Dx=\frac{dx}{\sqrt{2\pi}}e^{-\frac{x^{2}}{2}}$. \\
 \\
 Once we evaluate all of the integrals in the expression for $\langle V^{n}\rangle$
we can write it in the following form 
\begin{eqnarray}
\langle V^{n}\rangle=\exp(nN(G_{0}(q,\tilde{q},\tilde{r},m)+\alpha G_{1}(q,\tilde{q},\tilde{r},m)))
\end{eqnarray}
where $G_{0}(q,\tilde{q},m)$ is an entropic contribution coming from
from the integral over the weights and $G_{1}(q,\tilde{q},\tilde{r},m)$
is an energetic contribution whose form is dictated by the learning
rule. \\
 We can start by computing the energetic contribution. We define $A(t,\bar{h}_{0})$
and $Z(t,\bar{h}_{0})$ as 
\begin{eqnarray}
A(t,\bar{h}_{0}) & = & \frac{1}{2(r_{Q}-Q)}\left(\sqrt{Q-\bar{m}^{2}}t+\bar{h}_{0}\bar{m}\right)^{2}-\frac{\bar{m}^{2}\bar{h}_{0}^{2}}{2(r_{Q}-Q)}\\
Z(t,\bar{h}_{0}) & = & \int_{\kappa}^{\infty}\frac{dh}{\sqrt{2\pi}}\exp\left(-\frac{1}{2(r_{Q}-Q)}\left[h-\left(\sqrt{Q-\bar{m}^{2}}t+\bar{h}_{0}\bar{m}\right)\right]^{2}\right)
\end{eqnarray}
and rewrite $X^{\prime}$ as 
\begin{eqnarray}
X^{\prime} & = & 2\int_{0}^{\infty}D\bar{h}_{0}\int Dt\Big[\exp(nA(t,\bar{h}_{0}))Z^{n}(t,\bar{h}_{0})\Big]
\end{eqnarray}
We define $A$ as the average over $t$, $\bar{h}^{0}$ of $A(t,\bar{h}_{0})$, i.e. 
\begin{eqnarray}
	A & = & \int_{0}^{\infty}D\bar{h}^{0}\int DtA(t,\bar{h}_{0})=\frac{Q-\bar{m}^{2}}{2(r_{Q}-Q)}
\end{eqnarray}
In the limit $n\rightarrow0$, $X^{\prime}$ becomes
\begin{equation}
X^{\prime}=\exp\left(An+2n\int_{0}^{\infty}D\bar{h}^{0}\int Dt\log Z(t,\bar{h}^{0})\right)\label{eq:a2}
\end{equation}
We can do the following shift of variables 
\begin{align}
x & =\left(\sqrt{Q-\bar{m}^{2}}t+\bar{h}_{0}\bar{m}\right)/\sqrt{Q}\\
y & =\left(-\bar{m}t+\sqrt{Q-\bar{m}^{2}}\bar{h}_{0}\right)/\sqrt{Q}
\end{align}
which allows us to write $t$ and $\bar{h}^{0}$ as
\begin{align}
t & =\left(\sqrt{Q-\bar{m}^{2}}x-y\bar{m}\right)/\sqrt{Q}\\
\bar{h}_{0} & =\left(x\bar{m}+\sqrt{Q-\bar{m}^{2}}y\right)/\sqrt{Q}
\end{align}
and $Z(t,\bar{h}_{0})$ as 
\begin{eqnarray}
Z(x) & = & \int_{\kappa}^{\infty}\frac{dh}{\sqrt{2\pi}}\exp\left(-\frac{(h-\sqrt{Q}x)^{2}}{2(r_{Q}-Q)}\right)=\sqrt{r_{Q}-Q}H\left(\frac{\kappa-\sqrt{Q}x}{\sqrt{r_{Q}-Q}}\right)
\end{eqnarray}
Under this transformation, the Gaussian integrals become 
\begin{equation}
\int_{0}^{\infty}D\bar{h}^{0}\int Dt=\int Dx\int_{-x\bar{m}/\sqrt{Q-\bar{m}^{2}}}^{\infty}Dy=\int DxH\left(-x\bar{m}/\sqrt{Q-\bar{m}^{2}}\right)
\end{equation}
where we define
\begin{eqnarray}
H(x) & = & \int_{x}^{\infty}Dy
\end{eqnarray}
This gives us 
\begin{align}
2\int_{0}^{\infty}Dh^{0}\int Dt\log Z(t,h^{0})= & 2\int DxH\left(-x\bar{m}/\sqrt{Q-\bar{m}^{2}}\right)\log Z(x)\\
= & 2\int DxH\left(-x\bar{m}/\sqrt{Q-\bar{m}^{2}}\right)\log H\left(\frac{\kappa-\sqrt{Q}x}{\sqrt{r_{Q}-Q}}\right)+\frac{1}{2}\log(r_{Q}-Q)\nonumber 
\end{align}
So $X^{\prime}$ becomes 
\begin{eqnarray}
X^{\prime} & = & \exp\left(2\int DxH\left(-x\bar{m}/\sqrt{Q-\bar{m}^{2}}\right)\log H\left(\frac{\kappa-\sqrt{Q}x}{\sqrt{r_{Q}-Q}}\right)+\frac{1}{2}\log(r_{Q}-Q)+A\right)^{n}
\end{eqnarray}
Using the relation 
\begin{eqnarray}
A-\frac{1}{2n}\text{log}\text{det}(Q_{ab}-\bar{m}^{2})+\frac{1}{2}\log(r_{Q}-Q) & = & 0
\end{eqnarray}
the replicated volume of the version space become
\begin{align}
\langle V^{n}\rangle & =\int\prod_{a=1}^{n}d^{N_{0}}w^{a}d^{N_{1}}\tilde{w^{a}}\delta\left(\sum_{i=1}^{N_{0}}(w_{i}^{a})^{2}+\sum_{j=1}^{N_{+}}(\tilde{w}_{j}^{a})^{2}-N\right)\int dm\int dq_{ab}\int d\tilde{q}_{ab}\delta(Nm-\sum_{i=1}^{N_{0}}w_{i}^{a}w_{i}^{0})\nonumber \\
 & \prod_{ab}\delta\left(Nq_{ab}-\sum_{i=1}^{N_{0}}w_{i}^{a}w_{i}^{b}\right)\delta\left(N\tilde{q}_{ab}-\sigma_{in}^{2}\sum_{j=1}^{N_{+}}\tilde{w}_{j}^{a}\tilde{w}_{j}^{b}\right)\exp\Big(2n\int DxH\left(-\frac{x\bar{m}}{\sqrt{Q-\bar{m}^{2}}}\right)\log H\left(\frac{\kappa-\sqrt{Q}x}{\sqrt{r_{Q}-Q}}\right)\Big)^{P}\nonumber \\
\end{align}
We can compute the entropic term $G_{0}(q,\tilde{q},m,\tilde{r})$
by considering the integrals over configurations of weights allowed
by the delta functions. Then $\exp(nN(G_{0}(q,\tilde{q},\tilde{r},m))$
is given by 
\begin{eqnarray}
\exp(nN(G_{0}(q,\tilde{q},\tilde{r},m)) & = & \int\prod_{a=1}^{n}\int dw_{a}d\tilde{w}_{a}\delta\left(Nm-\sum_{i=1}^{N_{0}}w_{i}^{a}w_{i}^{0}\right)
\nn &\times&\prod_{ab}\delta\left(Nq_{ab}-\sum_{i=1}^{N_{0}}w_{i}^{a}w_{i}^{b}\right)\delta\left(N\tilde{q}_{ab}-\sigma_{in}^{2}\sum_{j=1}^{N_{+}}\tilde{w}_{j}^{a}\tilde{w}_{j}^{b}\right)\label{eq:G0}
\end{eqnarray}
Introducing the Lagrange multipliers $\hat{m}$, $\hat{q}_{ab}$,
and $\hat{\tilde{q}}_{ab}$, Eqn.~\ref{eq:G0} can be written as
\begin{align}
\exp(nN(G_{0}(q,\tilde{q},\tilde{r},m)) & =\int\prod_{a=1}^{n}dw^{a}d\tilde{w}^{a}\int\frac{d\hat{m}}{\sqrt{2\pi}}\int\frac{d\hat{q}_{ab}}{\sqrt{4\pi}}\int\frac{d\hat{\tilde{q}}_{ab}}{\sqrt{4\pi}}\exp\Big(\frac{i}{2}\sum_{ab}\hat{q}_{ab}\left(Nq_{ab}-\sum_{i=1}^{N_{0}}w_{i}^{a}w_{i}^{b}\right)\nonumber \\
 & +\frac{i}{2}\sum_{ab}\hat{\tilde{q}}_{ab}\left(N\tilde{q}_{ab}-\sigma_{in}^{2}\sum_{j=1}^{N_{+}}\tilde{w}_{j}^{a}\tilde{w}_{j}^{b}\right)+i\sum_{a}\hat{m}_{a}\left(Nm_{a}-\sum_{i=1}^{N_{0}}w_{i}^{a}w_{i}^{0}\right)\Big)\nonumber \\
 & =\int\frac{d\hat{m}}{\sqrt{2\pi}}\int\frac{d\hat{q}_{ab}}{\sqrt{4\pi}}\int\frac{d\hat{\tilde{q}}_{ab}}{\sqrt{4\pi}}\exp\left(\frac{iN}{2}\sum_{ab}\hat{q}_{ab}q_{ab}+\frac{iN}{2}\sum_{ab}\hat{\tilde{q}}_{ab}\tilde{q}_{ab}+iN\sum_{a}\hat{m}_{a}m_{a}\right)\nonumber \\
 & \times\int\prod_{a=1}^{n}dw^{a}d\tilde{w}^{a}\exp\left(-iH(w^{a},\tilde{w^{a}},w^{0})\right)
\end{align}
where we have defined a ``Hamiltonian'' $H(w_{a},\tilde{w}_{a},w_{0})$
as 
\begin{eqnarray}
H(w^{a},\tilde{w}^{a},w^{0}) & = & \frac{1}{2}\sum_{ab}(\hat{q}_{ab}\sum_{i=1}^{N_{0}}w_{i}^{a}w_{i}^{b}+\sigma_{in}^{2}\hat{\tilde{q}}_{ab}\sum_{j=1}^{N_{+}}\tilde{w}_{j}^{a}\tilde{w}_{j}^{b})+\sum_{a}\sum_{i=1}^{N_{0}}{w_{i}^{a}w_{i}^{0}\hat{m}_{a}}
\end{eqnarray}
Doing a Wick rotation $i\hat{m}_{a}\rightarrow\hat{m}_{a}$, $i\hat{q}_{ab}\rightarrow\hat{q}_{ab}$,
$i\hat{\tilde{q}}_{ab}\rightarrow\hat{\tilde{q}}_{ab}$ and integrating
over the weights $\vec{w}$ and $\tilde{\vec{w}}$, we have 
\begin{align}
\exp(nN(G_{0}(q,\tilde{q},\tilde{r},m) & =\int\frac{d\hat{m}}{\sqrt{2\pi}}\int\frac{d\hat{q}_{ab}}{\sqrt{4\pi}}\int\frac{d\hat{\tilde{q}}_{ab}}{\sqrt{4\pi}}\exp\left(\frac{N}{2}\sum_{ab}\hat{q}_{ab}q_{ab}+\frac{N}{2}\sum_{ab}\hat{\tilde{q}}_{ab}\tilde{q}_{ab}+N\sum_{a}\hat{m}_{a}m\right)\nonumber \\
 & \times\int\prod_{a=1}^{n}dw^{a}d\tilde{w}^{a}\exp\left(-\frac{1}{2}\sum_{ab}(\hat{q}_{ab}\sum_{i=1}^{N_{0}}w_{i}^{a}w_{i}^{b}+\sigma_{in}^{2}\hat{\tilde{q}}_{ab}\sum_{j=1}^{N_{+}}\tilde{w}_{j}^{a}\tilde{w}_{j}^{b})-\sum_{\alpha}{\sum_{i=1}^{N_{0}}w_{i}^{a}w_{i}^{0}\hat{m}_{a}})\right)\nonumber \\
 & =\int\frac{d\hat{m}}{\sqrt{2\pi}}\int\frac{d\hat{q}_{ab}}{\sqrt{4\pi}}\int\frac{d\hat{\tilde{q}}_{ab}}{\sqrt{4\pi}}\exp\left(\frac{N}{2}\sum_{ab}\hat{q}_{ab}q_{ab}+\frac{N}{2}\sum_{ab}\hat{\tilde{q}}_{ab}\tilde{q}_{ab}+N\sum_{a}\hat{m}_{a}m\right)\nonumber \\
 & \times\exp\left(\frac{N}{2}\sum_{ab}\hat{m}_{a}\hat{q}_{ab}^{-1}\hat{m}_{b}-\frac{N\beta^{-1}}{2}\log\det\hat{q}-\frac{N(1-\beta^{-1})}{2}\log\det\hat{\tilde{q}}\sigma_{in}^{2}\right)
\end{align}
We can evaluate the integral on the saddle point by solving for $\hat{m}_{a}$,
$\hat{q}_{ab}$ ,and $\hat{\tilde{q}}_{ab}$ using the three saddle
point equations
\begin{align}
0 & =\frac{N}{2}m_{\gamma}+\frac{N}{2}\sum_{b}(\hat{q}_{cb}^{-1})\hat{m}_{b}\label{eq:replicasaddle1}\\
0 & =-\frac{N}{2}\sum_{ab}\hat{m}_{a}(\hat{q}_{ac})^{-1}(\hat{q}_{bd})^{-1}\hat{m}_{b}-\frac{N\beta^{-1}}{2}(\hat{q}_{cd})^{-1}+\frac{N}{2}q_{cd}\label{eq:replicasaddle2}\\
0 & =-\frac{N(1-\beta^{-1})}{2}(\hat{\tilde{q}}_{cd})^{-1}+\frac{N}{2}\tilde{q}_{cd}\label{eq:replicassadle3}
\end{align}
We make the following replica symmetric ansatz for $\hat{q}_{\alpha\beta}$
and $\hat{\tilde{q}}_{\alpha\beta}$ 
\begin{align}
\hat{q}_{ab}=(\hat{q}_{0}-\hat{q}_{1})\delta_{ab}+\hat{q}_{1}\\
\hat{\tilde{q}}_{ab}=(\hat{\tilde{q}}_{0}-\hat{\tilde{q}}_{1})\delta_{ab}+\hat{\tilde{q}}_{1}
\end{align}
Inserting these expressions into Eqns.~\ref{eq:replicasaddle1},
\ref{eq:replicasaddle2}, and \ref{eq:replicassadle3} gives us the
following scalar equations 
\begin{align}
\frac{1}{\hat{q}_{0}-\hat{q}_{1}} & =\beta(1-\tilde{r}-q)\\
\hat{m} & =-\frac{m}{\beta(1-\tilde{r}-q)}\\
\hat{q}_{1} & =-\frac{q-m^{2}}{\beta(1-\tilde{r}-q)^{2}}\\
\frac{1}{\hat{\tilde{q}}_{0}-\hat{\tilde{q}}_{1}} & =\frac{\beta}{\beta-1}(\sigma_{in}^{2}\tilde{r}-\tilde{q})\\
\hat{\tilde{q}}_{1} & =-\frac{\beta-1}{\beta}\frac{\tilde{q}}{(\sigma_{in}^{2}\tilde{r}-\tilde{q})^{2}}
\end{align}
Solving for $\hat{m}$, $\hat{q}_{0}$, $\hat{q}_{1}$, $\hat{\tilde{q}}_{0}$,
and $\hat{\tilde{q}}_{1}$ we find
\begin{eqnarray}
G(q,\tilde{q},\tilde{r},m) & = & \frac{1}{2}\Bigg(1+\frac{q-m^{2}}{\beta(1-\tilde{r}-q)}+\frac{\beta-1}{\beta}\frac{\tilde{q}}{\sigma_{in}^{2}\tilde{r}-\tilde{q}}+\frac{1}{\beta}\log\left(\beta(1-\tilde{r}-q)\right)+\frac{\beta-1}{\beta}\log\left(\frac{\beta}{\beta-1}(\sigma_{in}^{2}\tilde{r}-\tilde{q})\right)\Bigg)
\end{eqnarray}
In summary, we have 
\begin{eqnarray}
\langle V^{n}\rangle=\exp nN(G_{0}(q,\tilde{q},\tilde{r},m)+\alpha G_{1}(q,\tilde{q},\tilde{r},m))
\end{eqnarray}
where $m$ is given by the saddle point value 
\begin{eqnarray}
G_{0}(q,\tilde{q},\tilde{r},m) & = & \frac{1}{2}\left(1+\frac{1}{\beta}\left(\frac{q-m^{2}}{(1-\tilde{r}-q)}+\log\beta\left(1-\tilde{r}-q\right)\right)+\frac{\beta-1}{\beta}\left(\frac{\tilde{q}}{\sigma_{in}^{2}\tilde{r}-\tilde{q}}+\log\frac{\beta\left(\sigma_{in}^{2}\tilde{r}-\tilde{q}\right)}{\beta-1}\right)\right)\\
G_{1}(q,\tilde{q},\tilde{r},m) & = & 2\int DxH\left(-\frac{x\bar{m}}{\sqrt{Q-\bar{m}^{2}}}\right)\log H\left(\frac{\kappa-\sqrt{Q}x}{\sqrt{r_{Q}-Q}}\right)
\end{eqnarray}
\section{Max-margin limit in mean field theory}
In the max margin limit the uniqueness of the solutions for $w$ and
$\tilde{w}$ imply 
\begin{equation}
q\rightarrow1-\tilde{r},\,\,\tilde{q}\rightarrow\sigma_{in}^{2}\tilde{r},\,\,Q\rightarrow r_{Q}
\end{equation}
In general, $q$ and $\tilde{q}$ approach their max margin values
at different rates. To account for this we define the scaling factors
$\lambda$ and $\tilde{\lambda}$ as
\begin{align}
\lambda & =\frac{r_{Q}-Q}{1-\tilde{r}-q}\\
\tilde{\lambda} & =\frac{r_{Q}-Q}{\sigma_{in}^{2}\tilde{r}-\tilde{q}}
\end{align}
where $\lambda^{-1}+\tilde{\lambda}^{-1}=1$. This allows us to rewrite
$G_{0}(q,\tilde{q},\tilde{r},m)$ so that all of the singular terms
scale as $(r_{Q}-Q)^{-1}$ as follows 
\begin{eqnarray}
G_{0}(q,\tilde{q},\tilde{r},m) & = & \frac{1}{2}\left(1+\frac{\lambda}{\beta}\frac{q-m^{2}}{r_{Q}-Q}+\frac{\tilde{\lambda}(\beta-1)}{\beta}\frac{\tilde{q}}{r_{Q}-Q}+\frac{1}{\beta}\log\left(\beta\lambda^{-1}\left(r_{Q}-Q\right)\right)+\frac{\beta-1}{\beta}\log\left(\frac{\beta\tilde{\lambda}^{-1}}{\beta-1}\left(r_{Q}-Q\right)\right)\right)\label{eq:lambdasaddle}
\end{eqnarray}
Taking the max margin limit followed by the limit $n\rightarrow0$,
we find that the free energy is given by 
\begin{eqnarray}
\langle\log V\rangle & = & \frac{N}{2(r_{Q}-Q)}\left(\frac{\lambda(1-\tilde{r}-m^{2})+\lambda(\lambda-1)^{-1}(\beta-1)\sigma_{in}^{2}\tilde{r}}{\beta}-2\alpha\int DxH\left(-\frac{x\bar{m}}{\sqrt{r_{Q}-\bar{m}^{2}}}\right)[\kappa-\sqrt{r_{Q}}x]_{+}^{2}\right)\label{eq:kappasaddle}
\end{eqnarray}
The saddle point equation for $m$ is 
\begin{eqnarray}
\frac{\lambda\bar{m}}{\sqrt{r_{Q}-\bar{m}^{2}}} & = & \frac{\alpha\beta}{\sqrt{2\pi}}\Bigg(\int_{-\frac{\kappa}{\sqrt{r_{Q}-\bar{m}^{2}}}}^{\infty}Dx\frac{x}{1+\sigma_{out}^{2}}\left(\frac{\kappa}{\sqrt{r_{Q}-\bar{m}^{2}}}+x\right)^{2}\Bigg)\label{eq:msaddle}
\end{eqnarray}
The saddle point equation for $\tilde{r}$ is 
\begin{eqnarray}
\frac{\lambda((\lambda-1)^{-1}(\beta-1)\sigma_{in}^{2}-1)}{\beta} & = & 2\alpha\int DxH\left(-\frac{x\bar{m}}{\sqrt{r_{Q}-\bar{m}^{2}}}\right)x(\kappa-\sqrt{r_{Q}}x)_{+}\frac{(\sigma_{in}^{2}-1)}{\sqrt{r_{Q}}}\label{eq:rsaddle}\\
 & + & \frac{\alpha\bar{m}(\sigma_{in}^{2}-1)}{\sqrt{2\pi}r_{Q}}\int_{-\frac{\kappa}{\sqrt{r_{Q}-\bar{m}^{2}}}}^{\infty}Dx\sqrt{r_{Q}-\bar{m}^{2}}x\left(\frac{\kappa}{\sqrt{r_{Q}-\bar{m}^{2}}}+x\right)^{2}
\end{eqnarray}
We can use Eqn.~\eqref{eq:msaddle} to further simplify this as
\begin{eqnarray}
\frac{\lambda((\lambda-1)^{-1}(\beta-1)\sigma_{in}^{2}-1)}{\beta} & = & 2\alpha\int DxH\left(-\frac{x\bar{m}}{\sqrt{r_{Q}-\bar{m}^{2}}}\right)x(\kappa-\sqrt{r_{Q}}x)_{+}\frac{(\sigma_{in}^{2}-1)}{\sqrt{r_{Q}}}+\frac{(\sigma_{in}^{2}-1)(\sigma_{out}^{2}+1)\lambda\bar{m}^{2}}{\beta r_{Q}}\label{eq:eqrtilde}
\end{eqnarray}
For $\lambda$, we have the saddlepoint equation 
\begin{eqnarray}
1-m^{2} & = & \tilde{r}\left(1-\frac{(\beta-1)\sigma_{in}^{2}}{(\lambda-1)^{2}}\right)
\end{eqnarray}
which has the relevant solution
\begin{eqnarray}
\lambda & = & 1+\sqrt{\frac{(\beta-1)\sigma_{in}^{2}\tilde{r}}{1-\tilde{r}-m^{2}}}
\end{eqnarray}

$R$, the cosine of the angle between student and teacher, can be
written in terms of $m$ and $\tilde{r}$ as

\begin{equation}
R=\frac{m}{\sqrt{1-\tilde{r}}}
\end{equation}

For $\sigma_{in}=1$, i.e. the variance of the augmented units matches
the variance of the original input, Eqns.\ \ref{eq:kappasaddle},
\ref{eq:msaddle}, \ref{eq:rsaddle}, and \ref{eq:lambdasaddle} simplify
considerably and are given
\begin{eqnarray}
\frac{1}{N}\langle\ln V\rangle & = & \frac{1}{2(1-q)}\left(1-m^{2}-2\alpha\int DxH\left(-\frac{x\bar{m}}{\sqrt{1-\bar{m}^{2}}}\right)[\kappa-x]_{+}^{2}\right)\label{eq:saddle1}\\
\bar{m} & = & \frac{\alpha\sqrt{1-\bar{m}^{2}}}{\sqrt{2\pi}}\Bigg(\int_{-\frac{\kappa}{\sqrt{1-\bar{m}^{2}}}}^{\infty}Dx\frac{x}{1+\sigma_{out}^{2}}\left(\frac{\kappa}{\sqrt{1-\bar{m}^{2}}}+x\right)^{2}\Bigg)\label{eq:saddle2}\\
\tilde{r} & = & \frac{\beta-1}{\beta}(1-m^{2})\label{eq:saddle3}\\
\lambda & = & \beta\label{eq:saddle4}\\
r_{Q} & = & 1
\end{eqnarray}

We can now write $R$ directly in terms of $m$ and $\beta$ as

\begin{align}
R & =\frac{m}{\sqrt{1-\frac{\beta-1}{\beta}(1-m^{2})}}\label{eq:rsigmain1}
\end{align}

\section{Network at capacity}

\label{app:capacity}

We determine the capacity of the network for fixed $\beta$ by setting
the margin $\kappa=0$ in the mean field equations. After performing
all of the integrals, we have the following three equations 
\begin{eqnarray}
\frac{\lambda(1-\tilde{r}-m^{2})+\lambda(\lambda-1)^{-1}(\beta-1)(\sigma_{in}^{2}\tilde{r})}{\beta} & = & \frac{\alpha}{\pi}\left(\text{arccot}\left(\frac{\bar{m}}{\sqrt{r_{Q}-\bar{m}^{2}}}\right)-\frac{\bar{m}\sqrt{r_{Q}-\bar{m}^{2}}}{r_{Q}}\right)\\
\frac{\lambda\bar{m}}{\sqrt{r_{Q}-\bar{m}^{2}}} & = & \frac{\alpha\beta}{\pi}\frac{1}{1+\sigma_{out}^{2}}\\
\frac{\lambda((\lambda-1)^{-1}(\beta-1)\sigma_{in}^{2}-1)}{\beta} & = & \frac{\alpha(\sigma_{in}^{2}-1)}{\sqrt{2\pi}\sqrt{r_{Q}}}\left(1-\frac{\bar{m}}{\sqrt{r_{Q}}}\right)+\frac{(\sigma_{in}^{2}-1)(\sigma_{out}^{2}+1)\lambda\bar{m}^{2}}{\beta r_{Q}}
\end{eqnarray}

We can express $\alpha$ as $\alpha=\alpha_{0}/\beta$ and solve
these equations numerically for $\alpha_{0}$ to determine $\alpha_{c}$ 

For $\sigma_{in}=1$, the equations for network capacity become
\begin{eqnarray}
1-m^{2} & = & \frac{\alpha}{\pi}\left(\text{arccot}\left(\frac{\bar{m}}{\sqrt{1-\bar{m}^{2}}}\right)-\bar{m}\sqrt{1-\bar{m}^{2}}\right)\\
\frac{\bar{m}}{\sqrt{1-\bar{m}^{2}}} & = & \frac{\alpha}{\pi}\frac{1}{1+\sigma_{out}^{2}}
\end{eqnarray}

Note that these equations depend on $\alpha$ but not on $\beta$.
This implies that for $\sigma_{in}=1$, $\alpha_{c}$ is only a function
of $\sigma_{out}$ . The capacity of a network of size $\beta$ then
obeys the simple scaling relation.

\begin{equation}
\alpha_{c}(\beta,\sigma_{out})=\beta\alpha_{c}(1,\sigma_{out})
\end{equation}

\section{Calculation of the generalization error}

\label{app:generalization}To evaluate the generalization error in
terms of the mean field order parameters, we start from the following
expression for the error 
\begin{eqnarray}
E(w,x,\epsilon) & = & \Theta\left(-\left(\frac{1}{\sqrt{N}}\sum_{i=1}^{N_{0}}w_{i}x_{i}\right)\left(\frac{1}{\sqrt{N}}\sum_{i=1}^{N_{0}}w_{i}^{0}\cdot x_{i}+\epsilon\right)\right)\label{eq:errorfunc}
\end{eqnarray}
Averaging over the input $\vec{x}$, and noise $\epsilon$, we get 

\begin{eqnarray}
E_{g}(w) & = & \int\prod_{i=1}^{N_{0}}\frac{dx_{i}}{\sqrt{2\pi}}e^{-\frac{x_{i}^{2}}{2}}\int\frac{d\epsilon}{\sqrt{2\pi\sigma_{out}^{2}}}e^{\frac{-\epsilon^{2}}{2\sigma_{out}^{2}}}\Theta\left(-\left(\frac{1}{\sqrt{N}}\sum_{i=1}^{N_{0}}w_{i}x_{i}\right)\left(\frac{1}{\sqrt{N}}\sum_{i=1}^{N_{0}}w_{i}^{0}\cdot x_{i}+\epsilon\right)\right)\\
 & = & \int\prod_{i=1}^{N_{0}}\frac{dx_{i}}{\sqrt{2\pi}}e^{-\frac{x_{i}^{2}}{2}}\int\frac{d\epsilon}{\sqrt{2\pi\sigma_{out}^{2}}}e^{\frac{-\epsilon^{2}}{2\sigma_{out}^{2}}}\int\frac{dh}{\sqrt{2\pi}}\int\frac{dh^{0}}{\sqrt{2\pi}}\int\frac{d\hat{h}}{\sqrt{2\pi}}\int\frac{d\hat{h}^{0}}{\sqrt{2\pi}}\Theta\left(-hh^{0}\right)\\
 & \times & \exp\left(-i\hat{h}h-i\hat{h}^{0}h^{0}+\frac{i}{\sqrt{N}}\sum_{i=1}^{N_{0}}(\hat{h}w_{i}x_{i}+\hat{h}^{0}w_{i}^{0}x_{i})+i\hat{h}^{0}\epsilon\right)\\
 & = & \int\frac{dh}{\sqrt{2\pi}}\int\frac{dh^{0}}{\sqrt{2\pi}}\int\frac{d\hat{h}}{\sqrt{2\pi}}\int\frac{d\hat{h}^{0}}{\sqrt{2\pi}}\Theta\left(-hh^{0}\right)\\
 & \times & \exp\left(-i\hat{h}h-i\hat{h}^{0}h^{0}-\frac{1}{2N}(\hat{h}^{2}\sum_{i=0}^{N_{0}}w_{i}^{2}+2\hat{h}\hat{h}^{0}\sum_{i=1}^{N_{0}}w_{i}^{0}w_{i}+(\hat{h}^{0})^{2}\sum_{i=0}^{N_{0}}w_{i}^{2})-\frac{\sigma_{out}^{2}}{2}(\hat{h}^{0})^{2}\right)
\end{eqnarray}

We set the normalization of the student and teacher to be 
\begin{align}
||w|| & =||w^{0}||=\sqrt{N}
\end{align}
and define the order parameter $R$ as the cosine of the angle between
teacher and student as 
\begin{eqnarray}
R & =\frac{1}{N}\sum_{i=1}^{N_{0}} & w_{i}w_{i}^{0}
\end{eqnarray}
After performing the integral over $\hat{h}^{0}$, we can define a
rescaled $R$ and $h^{0}$ as
\begin{align}
\bar{R} & =\frac{R}{\sqrt{1+\sigma_{out}^{2}}}\\
\bar{h}^{0} & =\frac{h^{0}}{\sqrt{1+\sigma_{out}^{2}}}
\end{align}
We can then perform the integral over $\hat{h}$ to get the following
integral over $h$ and $\bar{h}^{0}$

\begin{eqnarray}
E_{g}(R) & = & \int\frac{\mathrm{d}h}{\sqrt{2\pi}}\frac{\mathrm{d}h^{0}}{\sqrt{2\pi}}\frac{\mathrm{d}\hat{h}}{\sqrt{2\pi}}\Theta\left(-h\bar{h}^{0}\right)e^{-\frac{1}{2}(1-\bar{R}^{2})\hat{h}^{2}-i\hat{h}(h+\bar{h}^{0}\bar{R})-\frac{1}{2}(\bar{h}^{0})^{2}}\\
 & = & \int\mathrm{d}h\mathrm{d}\bar{h}^{0}\frac{1}{2\pi\sqrt{1-\bar{R}^{2}}}\Theta\left(-h\bar{h}^{0}\right)e^{-\frac{1}{2(1-\bar{R}^{2})}(h^{2}-2h\bar{h}^{0}\bar{R}+(\bar{h}^{0})^{2})}
\end{eqnarray}

This evaluates to 
\begin{eqnarray}
E_{g}(R) & = & \frac{1}{\pi}\left(\frac{\pi}{2}-\tan^{-1}\left(\frac{R}{\sqrt{1+\sigma_{out}^{2}-R^{2}}}\right)\right)\label{eq:fulleg}
\end{eqnarray}
In our expanded network, $m$ and $R$ are related as 
\begin{eqnarray}
m & = & \frac{1}{N}R\|w^{0}\|\|w\|
\end{eqnarray}
This gives us 
\begin{eqnarray}
R & = & \frac{m}{\sqrt{1-\tilde{r}}}\label{eq:rdef}
\end{eqnarray}
In terms of $m$ and $\tilde{r}$ this can be written as 
\begin{equation}
E_{g}(m,\tilde{r}) = \frac{1}{\pi}\left(\frac{\pi}{2}-\tan^{-1}\left(\frac{m}{\sqrt{(1-\tilde{r})(1+\sigma_{out}^{2})-m^{2}}}\right)\right)
\end{equation}
\section{Large $\beta$ limit}
We can find a closed expression for the generalization error in the
limit $\beta\rightarrow\infty$ with $\sigma_{in}\leq1$. In this
limit we have $m\ll1$, $\alpha_{0}\ll\beta$ and $1\ll\kappa$. Analysis
of the saddle point equations gives us the following relations

\begin{align}
\sigma_{in}^{2} & =\frac{\alpha_{0}}{\beta}\kappa^{2}\label{eq:largebeta1}\\
\sigma_{in}^{2}-\beta^{-1}\lambda & =0\\
\lambda\bar{m} & =\frac{2\alpha_{0}}{\sqrt{2\pi}}\kappa\\
\beta\sigma_{in}^{2}\bar{m} & =\frac{2\alpha_{0}}{\sqrt{2\pi}}\sigma_{in}\sqrt{\frac{\beta}{\alpha_{0}}}\label{eq:largebeta2}\\
\lambda & =\sigma_{in}\sqrt{\frac{\beta}{1-\tilde{r}-m^{2}}}
\end{align}

which lead to the following expressions for $m$ and $\tilde{r}$

\begin{align}
m & =\sqrt{\frac{2\alpha_{0}\left(1+\sigma_{out}^{2}\right)}{\beta\pi\sigma_{in}^{2}}}\\
1-\tilde{r} & =\frac{1}{\beta\sigma_{in}^{2}}\left(1+2\pi^{-1}\alpha_{0}\left(1+\sigma_{out}^{2}\right)\right)
\end{align}

Plugging these into Eqn.\ \ref{eq:rdef} gives us

\begin{align}
R^{2} & \thickapprox\frac{\frac{2\alpha_{0}}{\pi}\frac{1}{(1+\sigma_{out}^{2})}}{1+\frac{2\alpha_{0}}{\pi}\frac{1}{(1+\sigma_{out}^{2})}}\nonumber \\
 & \approx1-\frac{\pi(1+\sigma_{out}^{2})}{2\alpha_{0}}\label{eq:rsquare2}
\end{align}

The expression for $R^{2}$ in Eqn.\ \eqref{eq:rsquare2} can be
plugged into Eqn.\ \eqref{eq:fulleg} to find an expression for the
generalization error for $\beta\rightarrow\infty$ which is shown
in Fig.\ \eqref{fig:thryaa}. Note that this expression does not
depend on $\sigma_{in}$ as long as $\sigma_{in}\leq1$

\section{Optimal input noise}

\label{app:opt_input} We find the optimal $\sigma_{in}$ to minimize
the generalization error by maximizing $R$. Differentiating $R$
with respect to $\sigma_{in}$ gives us 
\begin{eqnarray}
\frac{dR}{d\sigma_{in}} & = & \frac{dm}{d\sigma_{in}}\frac{1}{\sqrt{1-\tilde{r}}}+\frac{1}{2}\frac{d\tilde{r}}{d\sigma_{in}}\frac{m}{(1-\tilde{r})^{\frac{3}{2}}}
\end{eqnarray}
which gives us the condition 
\begin{eqnarray}
\frac{dm}{d\sigma_{in}} & = & -\frac{1}{2}\frac{m}{(1-\tilde{r})}\frac{d\tilde{r}}{d\sigma_{in}}\label{eq:sigmainopt}
\end{eqnarray}
\end{document}